\documentclass[12pt,english,aps,prb,superscriptaddress,showpacs,showkeys]{revtex4-1}
\usepackage[T1]{fontenc}
\usepackage[latin9]{inputenc}
\usepackage{geometry}
\usepackage{xcolor}
\usepackage{amsbsy}
\usepackage{amstext}
\usepackage{graphicx}
\PassOptionsToPackage{normalem}{ulem}
\usepackage{ulem}

\makeatletter

\newcommand{\lyxmathsym}[1]{\ifmmode\begingroup\def\b@ld{bold}
  \text{\ifx\math@version\b@ld\bfseries\fi#1}\endgroup\else#1\fi}

\providecommand{\tabularnewline}{\\}
\providecolor{lyxadded}{rgb}{0,0,1}
\providecolor{lyxdeleted}{rgb}{1,0,0}

\DeclareRobustCommand{\lyxsout}[1]{\ifx\\#1\else\sout{#1}\fi}

\makeatother

\usepackage{babel}
\begin{document}

\title{Dislocation content of grain boundary phase junctions and its relation
to grain boundary excess properties}

\author{T. Frolov}

\address{Lawrence Livermore National Laboratory, Livermore, California 94550,
USA}

\author{D. L. Medlin}

\address{Sandia National Laboratories, Livermore, California 94551, USA}

\author{M. Asta }

\address{Department of Materials Science and Engineering, University of California
Berkeley, Berkeley, California 94720, USA}
\begin{abstract}
We analyze the dislocation content of grain boundary (GB) phase junctions,
\textit{i.e.}, line defects separating two different GB phases coexisting
on the same GB plane. While regular GB disconnections have been characterized
for a variety of interfaces, GB phase junctions formed by GBs with
different structures and different numbers of excess atoms have not
been previously studied. We apply a general Burgers circuit analysis
to calculate the Burgers vectors \textbf{$\mathbf{b}$} of junctions
in two $\Sigma5$ Cu boundaries previously simulated with molecular
dynamics. The Burgers vectors of these junctions cannot be described
by the displacement shift complete (DSC) lattice alone. We show that,
in general, the normal component of \textbf{$\mathbf{b}$} is not
equal to the difference in the GB excess volumes, but contains another
contribution from the numbers of GB atoms per unit area $\Delta N^{*}$
required to transform one GB phase into another. In the boundaries
studied, the latter component dominates and even changes the sign
of \textbf{$\mathbf{b}$}. We derive expressions for the normal and
tangential components of \textbf{b} in terms of the DSC lattice vectors
and the non-DSC part due to $\Delta N^{*}$ and additional GB excess
properties, including excess volume and shears. These expressions
provide a connection between GB phase transformations driven by the
GB free energy difference and the motion of GB junctions under applied
normal and shear stresses. The proposed analysis quantifies \textbf{b}
and therefore makes it possible to calculate the elastic part of the
energy of these defects, evaluate their contribution to the nucleation
barrier during GB phase transformations, and treat elastic interactions
with other defects. 
\end{abstract}
\maketitle

\section{Introduction}

Similarly to bulk materials, interfaces can exist in multiple states,
or phases, and exhibit first-order phase transitions.\citep{Cantwell20141}
Such transitions proceed by nucleation and growth of a new interfacial
structure resulting in a discontinuous change in the excess properties
of the interface.\citep{Frolov2013} For fluid interfacial phases,
the conditions of equilibrium and stability were first derived by
Gibbs, who called them interfacial states\citep{Gibbs} Gibbs showed
that an interface should transform to a state with the lowest interfacial
free energy. In order to describe such transitions within the existing
theories of nucleation and phase transformation, it is necessary to
quantify the driving force as well as the nucleation barrier. The
interfacial free energy difference provides the driving force according
to Gibbs, while the excess free energy associated with the line defect
separating the two phases is the penalty for the transformation. A
thermodynamic framework quantifying the excess properties of such
line defects and their free energy has been recently proposed.\citep{Frolov:2015ab}
The developed framework is Gibbsian in spirit and makes no assumptions
about the atomic details of the defect structure. It assumes that
the energy of this defect is finite, scales with its length, and is
independent of the system size. As a result, the treatment applies
to fluids and some solid systems,\citep{PhysRevB.95.155444} but cannot
be extended to general solid-solid interfaces without significant
approximations. 

A grain boundary is a solid-solid interface formed by two misoriented
crystals of the same material. The phase behavior of GBs has recently
become a topic of increased interest due to the accumulating experimental
and modeling evidence of phase transformations at such interfaces.\citep{Divinski2012,PhysRevLett.59.2887,Cantwell20141,krause_review_2019,Rickman2016225,doi:10.1080/095008399177020,Rohrer2016231,rupert_role_2016,Dillon20076208,OBrien2018,ABDELJAWAD2017528,rajeshwari_k_grain_2020,glienke_grain_2020}
The line defects formed between two different GB phases can have a
long range elastic field associated with them. GB disconnections are
examples of such line defects and have been analyzed for different
types of interfaces.\citep{HIRTH2013749,Hirth96} Disconnections on
GBs and phase boundaries have been investigated by modeling and experiments.\citep{Medlin2017383,medlin_accommodation_2003,medlin_accommodation_2006,Pond03a,hirth_disconnections_2016,rajabzadeh_role_2014,zhu_situ_2019}
On the other hand, GB phase junctions have received much less attention
since there have only been a few studies of GB structures with two
different connected interfacial phases. At present, the topological
nature of these defects and the magnitude of their Burgers vectors
are not well understood. Recent modeling and experimental studies
in elemental Cu suggest that the dislocation character of these defects
has a strong effect on the kinetics of GB phase transformations and
could explain the low transformation rates at room temperature.\citep{meiners_observations_2020}
The elastic field contribution to the energy could, in principle,
dominate the GB phase nucleation behavior, and play a role in the
interaction of the junctions with other GB disconnections, lattice
dislocations and point defects. 

Atomistic simulations have demonstrated first-order transitions in
$\Sigma5(210)[001]$ and $\Sigma5(310)[001]$ symmetric tilt GBs in
Cu.\citep{Frolov2013} The heterogeneous structures of these two $\Sigma5$
GBs, each containing two different phases, are shown in Fig.~\ref{eq:bcircuit}.
Both structures were obtained by molecular dynamic simulations at
T=800K with the boundary connected to an open surface. In these simulations
the surface acts as a nucleation site and a source of extra atoms
or vacancies necessary to grow the new GB phase. Because the cores
of these GB phases are composed of different number of atoms, atoms
must be inserted or removed to transform from one phase to the other.\citep{doi:10.1080/01418618308243118,doi:10.1080/01418618608242811,DUFFY84a,Phillpot1992,Phillpot1994,Alfthan06,Alfthan07,Chua:2010uq,Frolov2013,Demkowicz2015}
During the transformation these GB phases are separated by a GB phase
junction. The line direction of this defect is normal to the plane
of the image and it spans the periodic dimension of the simulation
block. The transformation was accompanied by the migration of the
GB phase junction, which required diffusion of extra atoms from the
surface. These two $\Sigma5$ boundaries provide a convenient model
system to analyze the topological character of GB phase junctions
and quantify their Burgers vectors. In this work, we show that GB
phase junctions have a dislocation character, just like regular disconnections,
but their Burgers vectors differ from crystallographic DSC vectors
since they include a contribution resulting from the difference in
the GB structures. We demonstrate a general Burgers circuit analysis
that quantifies the dislocation content of these defects. We interpret
it in terms of contributions from different excess properties, including
excess volume and shear, and the difference in the number of atoms
required to transform one GB phase into another.

\section{Methodology }

\subsection{Closure failure and Burgers circuit construction \label{subsec:Closure-failure}}

\begin{figure}
\includegraphics[width=0.6\textwidth]{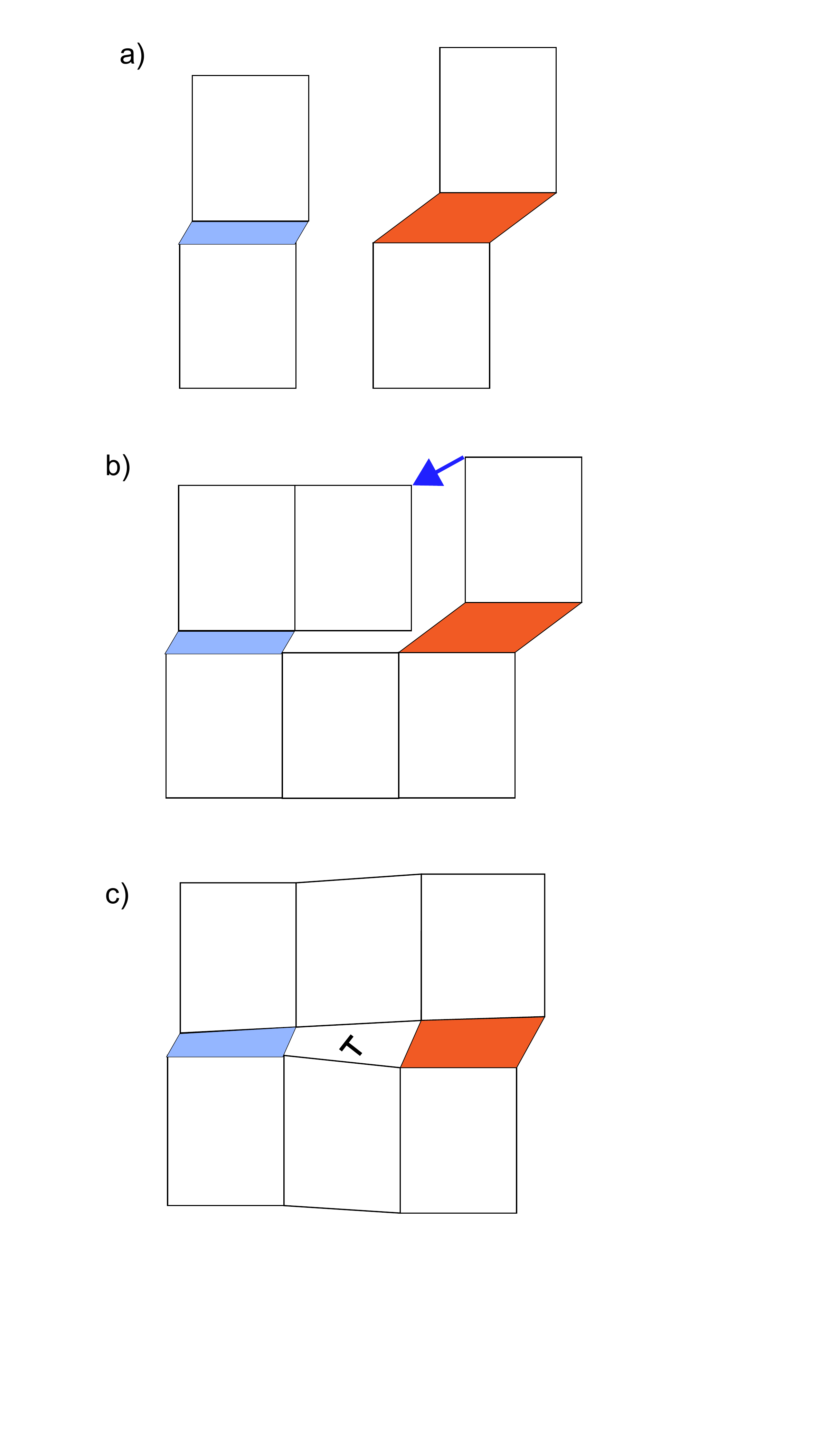}

\caption{Closure failure. a) Bicrystals with different GB structures, shown
in blue and orange, have different dimensions in the reference state.
b) An attempt to form a GB phase junction using the reference bicrystals
results in a closure failure. The bulk lattice planes cannot be joined
because they generally mismatch. c) GB phase junction is formed by
elastically deforming both bicrystals so the lattice planes in the
bulk can be connected. Both GB structures are elastically distorted
compared to a) and b). Here we assume that the system has a finite
size. In an infinitely large system, the blue and orange GB structures
would converge to their undistorted dimensions, shown in a) and b),
infinitely far away from the junction. In this construction, the magnitude
of the Burgers vector of the junction depends on the sizes and shapes
of the reference bicrystals shown in a). In general, different Burgers
vectors can be obtained between the same GB phases by changing the
bicrystals original dimensions. \label{fig:Closure_failure} }
\end{figure}

Fig. \ref{fig:Closure_failure} shows why GB phase junctions are expected
to have dislocation character and generate long-range elastic fields.
Consider two bicrystals with different GB structures or phases, as
shown schematically in Fig. \ref{fig:Closure_failure}a. Each GB structure
has its own excess properties such as excess volume\citep{Gibbs,Cahn79}
and shear.\citep{Frolov2012a,Frolov2012b} In addition, the GB cores
are composed of a different number of atoms. As a result the two bicrystals
are geometrically incompatible. To form a GB phase junction we need
to join the two bicrystals. Fig. \ref{fig:Closure_failure}b shows
that this procedure will result in a closure failure. In order to
form the junction using the Volterra operation,\citep{Hirth} we elastically
strain both bicrystals by appropriate amounts, so that their bulk
lattice planes away from the boundary match. As a result of this procedure,
we form a line defect with a dislocation character, as shown in Fig.
\ref{fig:Closure_failure}c. 

\begin{figure}
\includegraphics[width=0.7\textwidth]{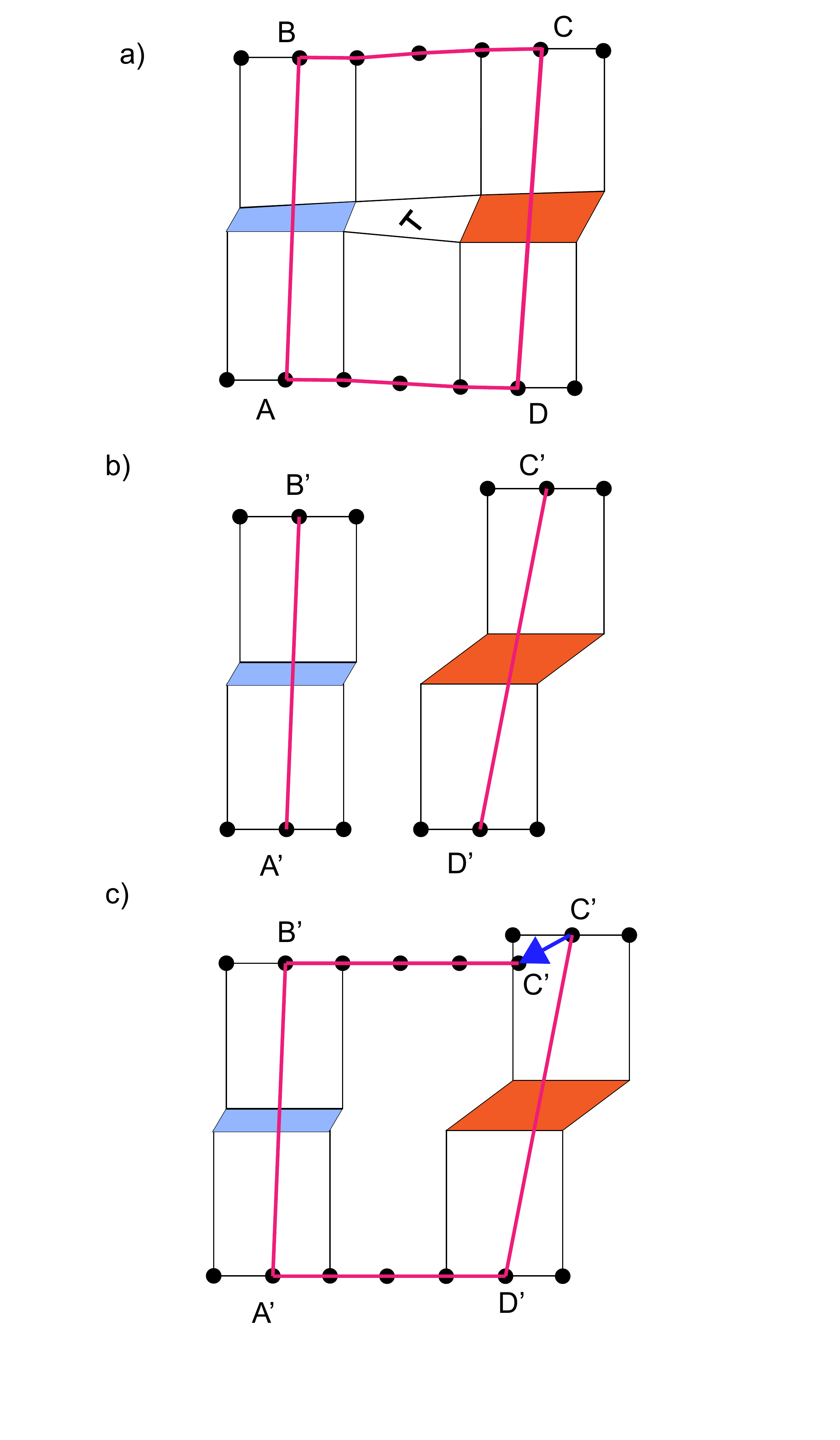}

\caption{Burgers circuit construction to calculate the Burgers vector of a
GB phase junction. a) A closed circuit is constructed around the GB
phase junction. b) The vectors crossing the different GB structures
are measured in the reference state. c) The closure failure in the
reference state gives the Burgers vector of the GB phase junction.
The unprimed and primed letters represent equivalent lattice sites
in the deformed and reference states, respectively. The black circles
represent lattice sites. \label{fig:Circuit} }
\end{figure}

Based on how the GB phase junction was created in this thought experiment,
it is straightforward to design a Burgers circuit that would quantify
the closure failure, \textit{i.e.}, the Burgers vector of the defect.
We start with the deformed bicrystal shown in Fig. \ref{fig:Circuit}
and construct a closed circuit ABCD around the GB phase junction.
The corner lattice points of the circuit coincide with the lattice
sites located inside the bulk crystal, far away from the GB junction
and GB phases. Two vectors $\mathbf{\mathbf{AB}}$ and $\mathbf{\mathbf{CD}}$
cross the two different GB phases, while the other two vectors $\mathbf{BC}$
and $\mathbf{DA}$ are located inside the bulk of the two grains and
are lattice vectors. To calculate the Burgers vector of the GB phase
junction we measure these vectors in the reference state before the
junction was created. The vectors $\mathbf{\mathbf{B^{\prime}C^{\prime}}}$
and $\mathbf{D^{\prime}A^{\prime}}$ are lattice vectors and their
components can be expressed in terms of the reference bulk lattice
parameter. The crossing vectors should be mapped on the respective
reference bicrystals as shown in Fig. \ref{fig:Circuit}b. Here, the
references for the crossing vectors are chosen as bicrystals possessing
only a single GB phase so that the crossing vectors will be unaffected
by the elastic field of the GB phase junction. With such a reference,
it is always possible to find a pair of of crystallographically equivalent
lattice sites, such as A' and B' and thereby measure the vector $\mathbf{A^{\prime}B^{\prime}}$.
Alternatively these equivalent lattice sites can be selected in the
same deformed bicrystal but infinitely far away from the junction.
The points A, B, C and D can be chosen arbitrary as long as they enclose
the phase junction and their positions in the reference state are
not affected by the GB. In other words, the difference between any
two choices of a given lattice point such as $A^{\prime}$ equals
a perfect lattice vector, which also means that the difference between
any two choices of the crossing vectors is a DSC vector. According
to the circuit construction, the closure failure, or the Burgers vector
$\mathbf{b}$, equals the sum of the four vectors measured in the
undeformed state:

\begin{equation}
\boldsymbol{\mathbf{b}}=\boldsymbol{A'B'}+\boldsymbol{B'C'}+\boldsymbol{C'D'}+\boldsymbol{D'A'}.\label{eq:bcircuit}
\end{equation}
Here, we follow start to finish right handed (SFRH) convention. In
Section \ref{sec: MD structures analysis}, we apply this approach
to quantify the dislocation content of GB phase junctions in the $\Sigma5(210)[001]$
and $\Sigma5(310)[001]$ Cu GBs. We describe how to calculate the
crossing vectors $\mathbf{A'B'}$ and $\mathbf{C'D'}$ in the reference
bicrystals. 

\subsection{Relating normal and tangential components of the Burgers vector to
excess properties of GBs and DSC lattice vectors. }

Equation (\ref{eq:bcircuit}) and the circuit procedure described
in Sec.~(\ref{subsec:Closure-failure}) are sufficient to calculate
the Burgers vector of any given GB phase junction.\footnote{In atomistic simulations this is always possible at least in principle.
The reference GB phases can be generated separately or carved out
from the deformed state and relaxed in a proper way so that the reference
crossing vectors could be calculated. In experiment, the analysis
cannot be completed based on a single image of the deformed state.
However, GB structures sufficiently far away from the junction can
be approximated as reference states. In general, a full three-dimensional
structure is necessary to determine all three components of $\mathbf{b}$.} The procedure, however, does not quantify the specific contributions
to $\mathbf{b}$ from bicrystallography and from the difference in
the GB excess properties. In this section, we express the crossing
vectors in terms of crystallographic properties and GB excess properties
and quantify their contributions to the Burgers vector. Relative normal
and tangential displacements of the grains due to the GB can be described
using the deformation gradient formalism.\citep{Frolov2012a,Frolov2012b}
The proposed thermodynamic framework enables quantification of the
excess shear at GBs and its relation to the relative translation of
the grains. Both the normal and tangential components of the crossing
vectors can be derived within this approach and related to the excess
volumes and shears of different GB phases. In the following two subsections
we will first derive the normal Burgers vector component $b_{\ensuremath{3}}$
using the well-known definition of the GB excess volume. In the next
section, we derive all three components of $\mathbf{\mathbf{b}}$
using the deformation gradient approach proposed in Refs.\citep{Frolov2012a,Frolov2012b}.

\subsubsection{Normal component $b_{\ensuremath{3}}$}

\begin{figure}
\includegraphics[width=1\textwidth]{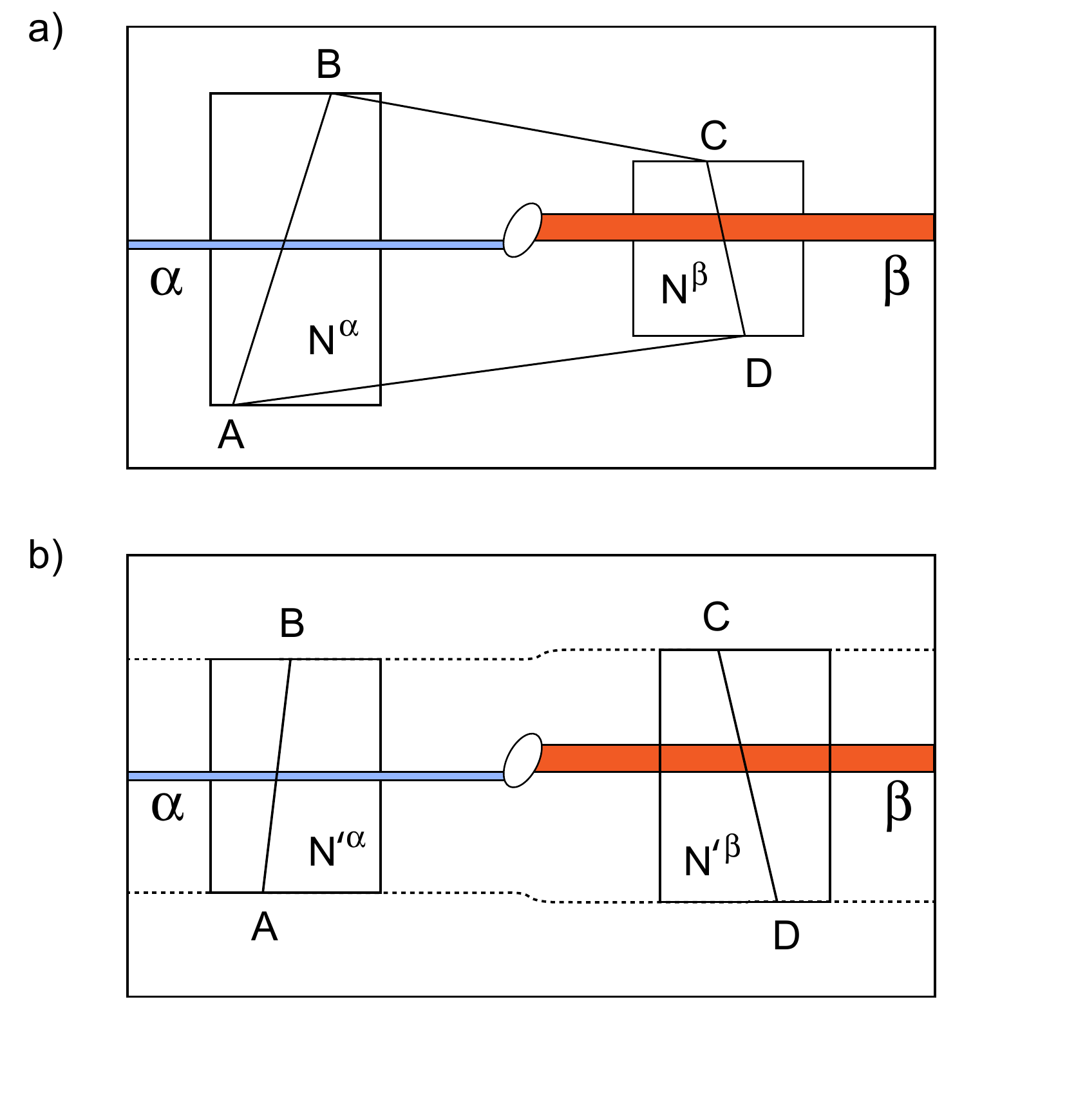}

\caption{a) A general Burgers circuit ABCD around a GB phase junction in the
deformed state. $\mathbf{AB}$ and $\mathbf{CD}$ are the crossing
vectors of the GB phases $\alpha$ and $\beta$, respectively. $\mathbf{BC}$
and $\mathbf{DA}$ are the lattice vectors of the upper and lower
grains, respectively. Their sum in the reference state is a DSC lattice
vector. $N^{\alpha}$ and $N^{\beta}$ are the numbers of atoms inside
the regions spanned by the crossing vectors. In this case the difference
$N^{\beta}-N^{\alpha}$ depends on the choice of the corners of the
circuit. b) A particular choice of the circuit around the same junction,
with the lattice sites B, C and D, A located on the same lattice planes
of the upper and lower crystals, respectively. These planes run parallel
to the GB plane and are indicated by the dashed lines. They are elastically
deformed due to the presence of the dislocation at the GB phase junction.
The quantity $\Delta N'/A=(N^{\prime\beta}-N^{\prime\alpha})/A$ is
the defect absorption capacity of this junction: it represents the
number of point defects per unit area which is absorbed or ejected
when the junction moves along the GB. \label{fig:Circuit-DAC} }
\end{figure}

The normal components of the crossing vectors measured in the reference
state can be expressed as a sum of bulk and GB contributions using
the familiar definition of the GB excess volume. The GB excess volume
$[V]_{N}$ is equal to the difference between the volume of the bicrystal
$V_{bicrystal}$ spanned by the crossing vector and containing $N$
atoms, and the volume of the bulk crystal containing the same number
of atoms divided by the GB area. The regions spanned by the crossing
vectors are shown in Fig. \ref{fig:Circuit-DAC}a. For two GB phases
$\alpha$ and $\beta$ we have

\begin{equation}
[V]_{N}^{\alpha}\cdot\mathcal{A}=V_{bicrystal}^{\alpha}-\Omega\cdot N^{\alpha}=A'B'_{3}\cdot A-\Omega\cdot N^{\alpha},\label{eq:exV_alpha}
\end{equation}

\begin{equation}
[V]_{N}^{\beta}\cdot\mathcal{A}=V_{bicrystal}^{\beta}-\Omega\cdot N^{\beta}=D'C'_{3}\cdot A-\Omega\cdot N^{\beta},\label{eq:exV_beta}
\end{equation}
where $\Omega$ is volume per atom in the perfect lattice, $\mathcal{A}$
is the GB area and the superscripts refer to bicrystals with GB phases
$\alpha$ and $\beta$. Eqs. (\ref{eq:exV_alpha}) and (\ref{eq:exV_beta})
express the normal components of the crossing vectors in the reference
state in terms of the excess volumes and numbers of atoms per unit
area:

\begin{equation}
A'B'_{3}=[V]_{N}^{\alpha}+\Omega\cdot N^{\alpha}/\mathcal{A},\label{eq:AB}
\end{equation}

\begin{equation}
D'C'_{3}=[V]_{N}^{\beta}+\Omega\cdot N^{\beta}/\mathcal{A}.\label{eq:CD}
\end{equation}
Here, both the GB excess volume and the second terms on the right-hand
side, $\Omega\cdot$$N/\mathcal{A}$, have the units of length. To
illustrate the physical meaning of the second term, consider first
a region inside a perfect crystal. In this case, $\Omega\cdot$$N/\mathcal{A}$
equals the distance between the corresponding atomic planes, which
is a normal component of a lattice vector. On a DSC lattice formed
by two perfect crystals with different orientations, this quantity
is equal to the normal component of a DSC vector. However, for a general
bicrystal with a GB, $\Omega\cdot$$N/\mathcal{A}$ is not necessarily
a component of a DSC vector because the number of atoms at the GB
is not restricted to be the same as in the lattice. We can now combine
Eqs (\ref{eq:AB}) and (\ref{eq:CD}) with Eq.\ (\ref{eq:bcircuit})
to derive an expression for $b_{3}$: 

\begin{equation}
b_{3}=(A'B'_{3}-D'C'_{3})+(B^{\prime}C_{3}^{\prime}-AD_{3}^{\prime})=[V]_{N}^{\alpha}-[V]_{N}^{\beta}+\Omega(N^{\alpha}-N^{\beta})/\mathcal{A}+d_{3}^{sc}=\Delta[V]_{N}+\Omega\Delta N/\mathcal{A}+d_{3}^{sc}.\label{eq:b_normal_analytic_alpha_beta}
\end{equation}
Equation\ (\ref{eq:b_normal_analytic_alpha_beta}) is an analytical
expression for the normal component of the Burgers vector of a GB
phase junction. The equation reveals the different factors that contribute
to $b_{3}$. The first one is the difference between the excess volumes
of the two GB phases. The second term corresponds to the difference
in the number of atoms inside the two regions spanned by the crossing
vectors. Finally, the two lattice vectors of the upper and lower grains
contribute the DSC component $d_{3}^{sc}=BC_{3}^{\prime}-AD{}_{3}^{\prime}$.

We derived Eq.\ (\ref{eq:b_normal_analytic_alpha_beta}) for one
particular GB phase junction, but we can use this analysis to demonstrate
that all other possible Burgers vectors of such junctions form a DSC
lattice. In addition, according to Eq.\ (\ref{eq:b_normal_analytic_alpha_beta})
the origin of this lattice is shifted away from zero in the direction
normal to the GB plane by $\Delta[V]_{N}+\Omega\Delta N/\mathcal{A}$.
To do so, we consider all other possible junctions between the same
GB phases with different Burgers vectors and construct the Burgers
circuits for each junction such that the crossing vectors $\mathbf{A'B'}$
and $\mathbf{C'D'}$ are the same for all junctions and $\mathbf{B'C^{\prime}}$
and $\mathbf{A'D^{\prime}}$ may not be the same. Then the difference
between any two Burgers vectors will be equal to the difference in
$\mathbf{B'C^{\prime}}-\mathbf{A'D^{\prime}}$ of the two circuits,
which is necessarily a DSC vector since $\mathbf{B'C^{\prime}}$ and
$\mathbf{A'D^{\prime}}$ are lattice vectors of different grains.
As a result the Burgers vectors of any two junctions differ exactly
by some DSC vector and all admissible Burgers vectors form a DSC lattice.

So far we assumed that the corners of the circuit were chosen in a
general way, as shown in Fig. \ref{fig:Circuit-DAC}a. We can now
consider some particular choices of the lattice sites A, B, C and
D to relate the terms in Eq.\ (\ref{eq:b_normal_analytic_alpha_beta})
to some measurable properties of GB junctions. Specifically, consider
a circuit shown in Fig. \ref{fig:Circuit-DAC}b when the lattice vectors
$\mathbf{BC}$ and $\mathbf{AD}$ are located along atomic planes
parallel to the GB plane. These atomic planes are elastically distorted
due to the presence of the dislocation, but they are parallel to each
other and to the GB plane in the reference state. By this choice,
$B^{\prime}C_{3}^{\prime}\equiv A^{\prime}D_{3}^{\prime}\equiv0$
setting the DSC term in Eq. (\ref{eq:b_normal_analytic_alpha_beta})
to zero and we obtain

\begin{equation}
b_{3}=\Delta[V]_{N}+\Omega\Delta N'/\mathcal{A}.\label{eq:b_normal_analytic_alpha_beta-DAC}
\end{equation}
Here, we define $\Delta N'/\mathcal{A}$ as the defect absorption
capacity of a GB junction. It is equal to the difference in the number
atoms inside the equivalent volumes located on the two sides of the
junction and bound by the same atomic planes parallel to the GB plane
shown in Fig.\ \ref{fig:Circuit-DAC}b. The defect absorption capacity
represents the number of atoms per unit of the GB area absorbed or
ejected when the junction moves along the GB. 

As a simple illustration, consider the climb of a regular disconnection
inside a single-phase GB. The motion of this disconnection requires
a supply of atoms or vacancies with the number of the point defects
proportional to the normal component of the Burgers vector, which
in this case is a DSC lattice vector.\citep{HIRTH2013749,Hirth96}
According to Eq.\ (\ref{eq:b_normal_analytic_alpha_beta-DAC}) the
defect absorption capacity of such a disconnection is given by $\Delta N_{disc}'/\mathcal{A}=b_{3}/\Omega=d_{3}^{sc}/\Omega$,
since the GB structure on the two sides of the disconnection is the
same and $\Delta[V]_{N}\equiv0$. This example also illustrates that
different disconnections may have different defect absorption capacities
and the difference is given by a normal component of some DSC vector
divided by the volume per atom. For a general GB phase junction separating
different GB phases, however, the defect absorption capacity is not
defined by the DSC lattice alone, as will be discussed below. 

For a given physical system, $b_{3}$ of a GB phase junction is a
well defined, single-valued quantity. However, any GB phase junction
can in principle increase or decrease its dislocation content and
its defect absorption capacity by absorbing or ejecting regular disconnections.
This multiplicity of possible $b_{3}$ is captured by the second term
in Eq.\ (\ref{eq:b_normal_analytic_alpha_beta-DAC}). While the first
term, $\Delta[V]_{N}=[V]_{N}^{\alpha}-[V]_{N}^{\beta}$, is a constant
set by the values of the excess volumes of the two GB phases, the
second term, $\Omega\Delta N'/\mathcal{A}=\Omega(N^{\alpha}-N^{\beta})/\mathcal{A}$,
represents a set of possible values. When disconnections are ejected
or absorbed by the junction, $\Omega\Delta N'/\mathcal{A}$ term describes
the discrete changes in the normal component of the Burgers vector
which occurs in increments dictated by the DSC lattice. For this reason,
it makes sense to further split $\Omega\Delta N'/\mathcal{A}$ in
Eq.\ (\ref{eq:b_normal_analytic_alpha_beta-DAC}) into contributions
described by the DSC lattice and the smallest in magnitude non-DSC
part $\Delta N^{*}/\mathcal{A}$ :

\begin{equation}
b_{3}=\Delta[V]_{N}+\Omega\Delta N^{*}/\mathcal{A}+d_{3}^{sc}\label{eq:b_normal_alytic_V_N_dSC}
\end{equation}
Subtracting Eqs.\ (\ref{eq:b_normal_analytic_alpha_beta-DAC}) and
(\ref{eq:b_normal_alytic_V_N_dSC}) we also obtain

\begin{equation}
\Delta N'/\mathcal{A}=\Delta N^{*}/\mathcal{A}+d_{3}^{sc}/\Omega\label{eq: DAC and DSC}
\end{equation}
Equation\ (\ref{eq:b_normal_alytic_V_N_dSC}) demonstrates that all
admissible Burgers vectors of GB phase junctions can be obtained by
constructing a DSC lattice for a given bicrystal and shifting the
origin of this lattice by a non-DSC vector with the normal component
given by $\Delta[V]_{N}+\Omega\Delta N^{*}/\mathcal{A}$. The in-plane
components of this shift vector will be derived in the next section.
Eq.\ (\ref{eq: DAC and DSC}) shows that the defect absorption capacity
of a junction can change in increments dictated by the DSC lattice,
but may not be reduced to zero in some cases because of $\Delta N^{*}$.
As introduced by Eq\ (\ref{eq:b_normal_alytic_V_N_dSC}), $\Omega\Delta N^{*}/\mathcal{A}$
is smaller than the smallest normal component of a DSC vector $min(d_{3}^{sc})$.
It is also defined up to the $min(d_{3}^{sc})$ and can be both positive
and negative.

To illustrate the meaning of the different terms in Eqs.\ (\ref{eq:b_normal_alytic_V_N_dSC})\ and\ (\ref{eq: DAC and DSC})
we consider several examples. We start again with a regular disconnection
as a simplest case when the GB structure is the same on both sides
of the line defect. The Burgers vector is exactly a DSC lattice vector,
as a result $\Delta N^{*}/\mathcal{A}\equiv0$. In other words, the
defect adsorption capacity of a regular disconnection is described
exactly by the normal components of the DSC lattice vectors. 

In another special case, consider a junction formed by two different
GB phases composed of the same number of atoms, meaning that given
a bicrystal with one GB phase, the same bicrystal with a different
GB phase can be obtained by rearranging the atoms in the GB region
and changing the relative translations of the grains if necessary
without inserting or removing atoms from the GB core. In this case
$\Delta N^{*}=0$ again, but the excess GB volume difference contributes
to the non-DSC part of the Burgers vector normal component: $b_{3}=[V]_{N}^{\alpha}-[V]_{N}^{\beta}+d_{3}^{sc}$.
As a result, differently from regular disconnections, the origin of
the DSC lattice of all possible Burgers vectors of such a junction
is not located at zero. In a general case, however, $\Omega\Delta N'/\mathcal{A}$
term is not equal to a normal component of a DSC vector or zero and
$\Delta N^{*}$ is not zero. 

At this point $\Delta N^{*}$ was derived through its contribution
to the normal shift of the origin of the DSC lattice of all possible
Burgers vectors of a given GB phase junction. We now turn our discussion
to the physical meaning of this quantity. We show that $\Delta N^{*}/\mathcal{A}$
corresponds to the smallest number of atoms or vacancies per unit
area required to transform one GB phase into another. Indeed, out
of all possible choices, we can always select a junction such that
the difference $b_{3}-\Delta[V]_{N}$ is smaller than the smallest
normal component $min(d_{3}^{sc})$ of any DSC vector, making $d_{3}^{sc}$
in Eq\ (\ref{eq:b_normal_alytic_V_N_dSC}) zero and $b_{3}-\Delta[V]_{N}=\Omega\Delta N^{*}/\mathcal{A}$.
For this junction, by definition, $\Delta N^{*}$ is the difference
in the number of atoms inside two regions containing two different
GB phases and can be interpreted as the number of atoms required to
be inserted or removed to transform one GB phase into another. This
number is also the smallest, because any other changes in the number
of atoms that preserve the two given GB structures require insertion
or removal of atoms in the increments of $min(d_{3}^{sc})/\Omega$
atoms per unit area. 

A growing number of modeling studies demonstrated that for many GB
transitions $\Delta N^{*}$ is not zero, some number of atoms must
be added or removed to transform one GB phase to the other. The difference
in the number of GB atoms $\Delta N^{*}$ originates from the fact
that some GB structures cannot be obtained by simply joining two perfect
half crystals: in addition, some fraction of the GB atoms needs to
be added or removed from the GB core and this fraction is different
for different GB phases. The importance of optimizing the number of
atoms at GBs has been demonstrated in different GB types and several
different materials systems.\citep{doi:10.1080/01418618308243118,doi:10.1080/01418618608242811,DUFFY84a,Phillpot1992,Phillpot1994,Alfthan06,Alfthan07,Chua:2010uq,Frolov2013,Demkowicz2015,Han:2016aa}
New computational methods designed to perform grand-canonical optimization
of GB structure have been proposed.\citep{Zhu2018,banadaki_efficient_2018,gao_interface_2019,yang_grain_2020} 

The optimization of the number of atoms in the GB core is related
to the atomic density at the boundary, but it is not uniquely determined
by the excess GB volume and represents an additional GB parameter.
Previous studies reported the actual number of removed or inserted
atoms for a given GB cross-section or calculated it per unit area
relative to an idealized reference bicrystal system which is arbitrary.\citep{Alfthan06,Demkowicz2015,Han:2016aa}
In our previous study, we reported a fraction of GB atoms $[n]$ calculated
relative to the numbers of atoms in a bulk plane parallel to the boundary.\citep{Frolov2013}
To calculate this quantity for a given GB, we count the total number
of atoms inside a region containing a GB and the number of atoms in
one atomic plane parallel to the GB located inside the perfect crystal.
The fraction $[n]$ was then calculated as a modulo of this two numbers
and was divided by the number of atoms in one plane. The advantage
of the quantity introduced in Ref.\ \citep{Frolov2013} is that it
allows to calculate a well defined property related to the numbers
of atoms at GBs without keeping track of the number of atoms inserted
or removed during the process of GB optimization. While this parameter
can be readily calculated for twist and symmetric tilt boundaries
for some crystal lattices, it cannot be accepted as a general descriptor.
For example, this quantity cannot be calculated for asymmetric boundaries
with different areal number density of atoms per plane in the different
grains. Moreover, even for symmetric boundaries this descriptor needs
to be generalized to work for crystal lattices with more than one
atom per primitive unit cell, such as in diamond or hexagonal close
packed (hcp) lattices. Note that for symmetric tilt GBs the number
of atoms per unit area in one bulk plane is given by $min(d_{3}^{sc})/\Omega$.
As a result, for such boundaries the proposed fraction of atoms $[n]$
is exactly equivalent to subtracting out the largest DSC component
and $[V]_{N}$ from the normal component of a crossing vector of a
given boundary and dividing it by $min(d_{3}^{sc})$, which is analogous
to Eq.\ (\ref{eq: DAC and DSC}) derived for $\Delta N^{*}$. As
a modulo, $[n]$ is required to be positive. The advantage of calculating
the smallest non-DSC component of a crossing vector derived in this
study instead of $[n]$ is that it is aslo defined for asymmetric
GBs. This non-DSC component can be calculated for each individual
boundary and is related to the excess number of GB atoms per unit
area which we denote as $N^{*}/\mathcal{A}$ relative to the bulk
system defined by the subtracted DSC vector component. By this definition,
this number of GB atoms per unit area is defined up to $min(d_{3}^{sc})/\Omega$
and can be positive and negative.

In the context of GB phase transformations analyzed in this work,
$\Delta N^{*}/\mathcal{A}$, representing the smallest number of atoms
or vacancies per unit of GB area required to transform one GB phase
into another, is a well defined quantity which can be measured for
symmetric and asymmetric GBs. The derived Eq. (\ref{eq:b_normal_alytic_V_N_dSC})
relates $\Delta N^{*}/\mathcal{A}$ to the non-DSC part of the normal
Burgers vector component of the GB phase junction. In our derivation
leading to Eq. (\ref{eq:b_normal_alytic_V_N_dSC}), we made no assumptions
about the type of the CSL boundary and it is valid for both symmetric
and asymmetric boundaries. We analyze specific examples of GB phase
junctions and calculate $\Delta N^{*}$ in the atomistic simulation
section of this article.

\subsubsection{Deformation gradient treatment of all three components of the Burgers
vector }

In the previous section we showed that the normal components of all
possible Burgers vectors of a GB phase junction can be described by
a DSC lattice with the origin shifted normal to the GB plane by a
vector related to the difference in the excess volumes and the number
of atoms $\Delta N^{*}$. Since a GB phase junction is a dislocation,
it will experience a Peach-Koehler (PK) force when mechanical stresses
are applied.\citep{Hirth} This force produces a driving force for
the junction motion \textit{i.e.}, GB phase transformation, and will
also influence the equilibrium coexistence. When tension or compression
is applied normal to the GB plane, the driving force or the work of
the PK force per unit area is equal to the product of the normal component
of stress and the normal component of the Burgers vector. 

Another way to describe the same effects is to consider the free energies
of the two phases. Consider a junction between two GB phases with
$\Delta N^{*}=0$. The mechanical stresses normal to the GB plane
change the free energies of both grain boundary phases, with the change
proportional to the excess volume of each boundary, as described by
the absorption equation \citep{Gibbs}. This part of the free energy
difference due to the normal stress contributes to the driving force
for the GB phase transformation and is given by the product of the
normal component of stress and the difference in the excess volumes.
So far, we have demonstrated that the normal component of the Burgers
vector contains a contribution from the difference in the excess volumes.
Thus, the analysis presented here for the normal component of $\mathbf{b}$
provides a connection between these two equivalent descriptions of
the driving force.

In addition to the normal stress, solid interfaces support shear stresses
parallel to the interface plane.\citep{Robin74,dingreville_interfacial_2008,Larche_Cahn_78,doi:10.1063/1.448644}
These stresses also result in a PK force on GB phase junction and
change the free energies of the two phases. Excess shear of an interface
is an extensive property that describes how the interface free energy
changes with applied shear stress parallel to the boundary.\citep{Frolov2012a}
Excess shears and GB free energy as a function of shear stress have
been calculated for different GBs using atomistic simulations.\citep{Frolov2012b,meiners_observations_2020}
A recent study demonstrated a shear stress induced GB transformations
as well as equilibrium coexistence under applied shear stress. Moreover,
the coexistence stress was accurately predicted from the values of
the excess shears and the stress dependence of the GB free energies.\citep{meiners_observations_2020}

In this section we derive an expression for all three components of
$\mathbf{b}$, including the tangential components. We show that the
origin of this DSC lattice of possible Burgers vectors is also shifted
in the plane of the boundary due to the difference in excess shears.
To do so, we need to express all three components of the crossing
vectors in terms of contributions from the bulk and GB properties,
such as the GB excess volume, shear and the number of atoms. Following
Ref.\ \citep{Frolov2012a,Frolov2012b}, we assumed that there exists
a mapping of one grain into the other, which establishes a unique
relation between the lattice sites of the two crystals. This transformation
is described by a deformation gradient $\mathbf{F}$. In this work,
we only consider mappings that exclude transformations resulting in
GB coupled motion, which means that $\mathbf{\mathbf{F_{i3}^{b}}}$
components of the deformation gradient are the same for both grains,
where the superscript b indicated the bulk part of the crystals. Specific
examples of excess shear calculations for different GBs can be found
in Refs.\ \citep{Frolov2012b,meiners_observations_2020}.

\begin{figure}
\includegraphics[width=0.7\textwidth]{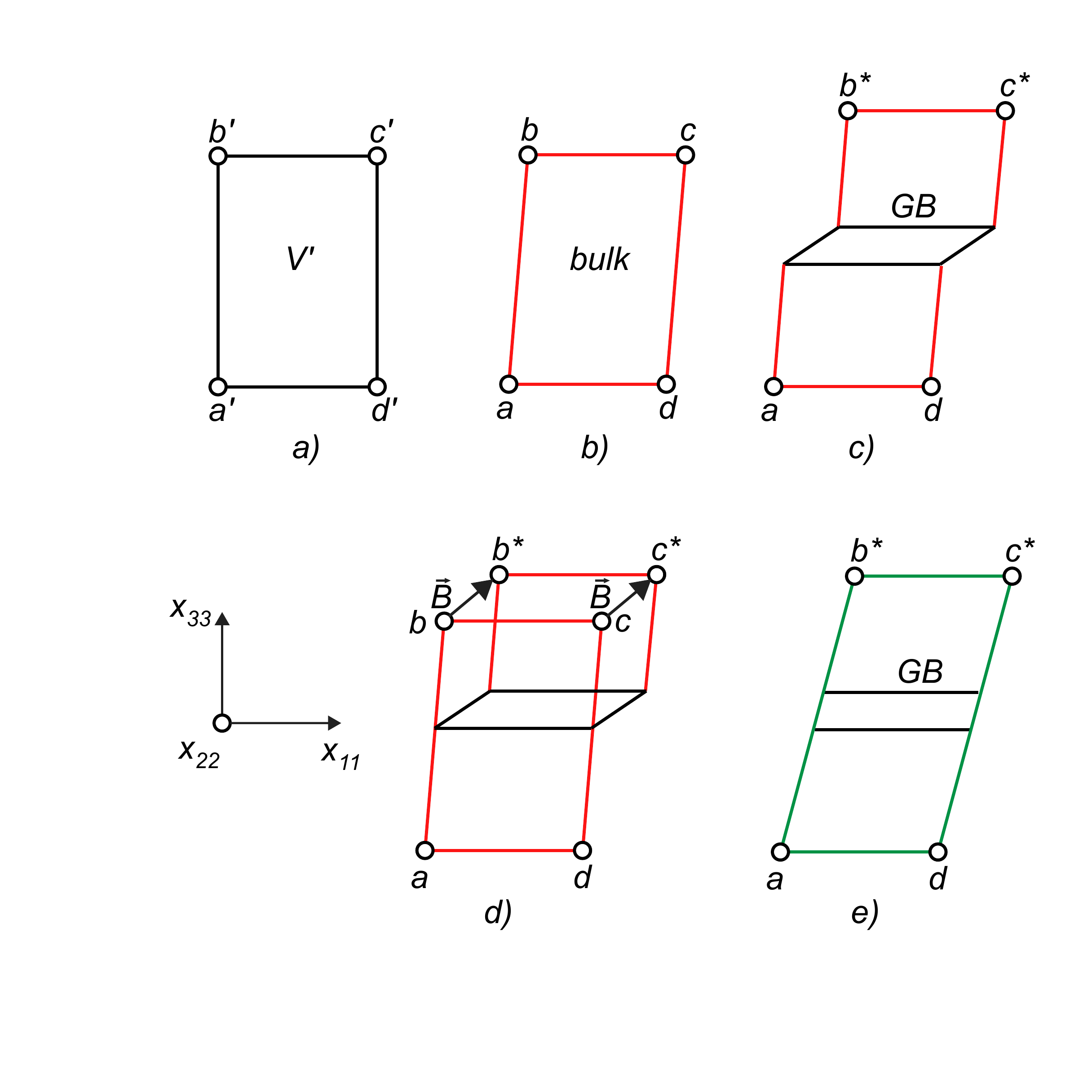}

\caption{Two-dimensional schematic of a mapping of a region of single-crystalline
material to a bicrystal containing a GB (reproduced from Ref.\ \citep{Frolov2012a}).
a) Reference state used to calculate the deformation gradient. b)
Actual, deformed state of the single crystal c) Region containing
the GB obtained from the single-crystalline region. d) Superimposed
single-crystal and bicrystal regions showing the displacement vector
$\mathbf{B}$. The open circles represent lattice sites labeled a
through d with the prime indicating the reference state and the asterisk
indicating the bicrystal. The parallelogram defined by the vertices
$a$,$b$, $c^{*}$ and $d^{*}$ is shown in e). Its mapping on the
reference state in a) defines the deformation gradient $\bar{\mathbf{F}}^{\alpha}$
producing the bicrystal region with a given GB phase $\alpha$. \label{fig:Two-dimentional-schematic_F_bulk_F_alpha}}
\end{figure}

The two-dimensional schematic in Fig.\ \ref{fig:Two-dimentional-schematic_F_bulk_F_alpha}
shows how the deformation gradient $\mathbf{F}$ can be used to describe
the mapping between a single crystal and a bicrystal with a GB phase.
Here we present equations for one phase ($\alpha$) and subsequently
distinguish GB phase specific properties by respective super-scripts.
We assume that there are lattice sites, or markers, that can be selected
and tracked during an imaginary transformation of a single crystal
in Fig. \ref{fig:Two-dimentional-schematic_F_bulk_F_alpha}b into
a bicrystal shown in Fig.\ \ref{fig:Two-dimentional-schematic_F_bulk_F_alpha}c.
These lattice sites, labeled as \textit{a}, \textit{b}, \textit{c}
and \textit{d} define a parallelogram (parallelepiped in three dimensions)
which has different shapes for the sites located in the single crystal
and the bicrystal. These shapes, shown in Figs\  \ref{fig:Two-dimentional-schematic_F_bulk_F_alpha}b
and e, can be described using deformation gradients $\mathbf{F}^{b}$
and $\bar{\mathbf{F}^{\alpha}}$ that map a common mathematical reference
state shown in Fig.\ \ref{fig:Two-dimentional-schematic_F_bulk_F_alpha}a
on both parallelograms. This mathematical reference state is used
to calculate the deformation gradients only and should not be confused
with the reference state for the Burgers circuit analysis discussed
earlier. Notice that, to describe the transformation from Fig.\ \ref{fig:Two-dimentional-schematic_F_bulk_F_alpha}
b to c, we use an effective deformation gradient $\bar{\mathbf{F}^{\alpha}}$
which is calculated based on the positions of the lattice sites as
shown in Fig.\ \ref{fig:Two-dimentional-schematic_F_bulk_F_alpha}
e. For three-dimensional systems, these deformation gradients are
given by\citep{Frolov2012a,Frolov2012b}

\[
F^{b}=\left(\begin{array}{ccc}
F_{11}^{b} & F_{12}^{b} & F_{13}^{b}\\
0 & F_{22}^{b} & F_{23}^{b}\\
0 & 0 & F_{33}^{b}
\end{array}\right)
\]

\begin{equation}
\bar{F}^{\alpha}=\left(\begin{array}{ccc}
F_{11}^{\alpha} & F_{12}^{\alpha} & \bar{F}_{13}^{\alpha}\\
0 & F_{22}^{\alpha} & \bar{F}_{23}^{\alpha}\\
0 & 0 & \bar{F}_{33}^{\alpha}
\end{array}\right)=\left(\begin{array}{ccc}
F_{11}^{b} & F_{12}^{b} & \left(F_{13}^{b}+B_{1}^{\alpha}\mathcal{A}^{'}/V^{'}\right)\\
0 & F_{22}^{b} & \left(F_{23}^{b}+B_{2}^{\alpha}\mathcal{A}^{'}/V^{'}\right)\\
0 & 0 & \left(F_{33}^{b}+B_{3}^{\alpha}\mathcal{A}^{'}/V^{'}\right)
\end{array}\right)\label{eq:F_alpha_B}
\end{equation}
where vector $\mathbf{B}^{\alpha}$ shown in Fig. \ref{fig:Two-dimentional-schematic_F_bulk_F_alpha}
d describes the change in the position of the site b relative to its
position in the single crystal. The coordinate axes are indicated
in the figure. From the formal definition of $\bar{\mathbf{F}}^{\alpha}$
by Eq.\ (\ref{eq:F_alpha_B}) and Figs\ \ref{fig:Two-dimentional-schematic_F_bulk_F_alpha}c
and e, it is clear that its $\bar{F}_{i3}^{\alpha}$ components depend
on the size of the selected GB region and approach bulk values when
the GB area to volume ratio $\mathcal{A}^{\prime}/V^{\prime}$ decreases.
We now recognize that $ab^{*}$ is a crossing vector and its components
can be expressed by $ab_{i}^{*}$=$V/\mathcal{A}\bar{F}_{i3}^{\alpha}/\bar{F}_{33}^{\alpha}$,
where $i=1,2,3$. There are three excess properties associated with
vector $\mathbf{B}^{\alpha}$: two GB excess shears $[VF_{13}/F_{33}]_{N}^{\alpha}$
and $[VF_{23}/F_{33}]_{N}^{\alpha}$ and one GB excess volume $[VF_{33}/F_{33}]_{N}^{\alpha}=[V]_{N}^{\alpha}$,
which can be found from the relation\citep{Frolov2012a,Frolov2012b}

\begin{equation}
[VF_{i3}/F_{33}]_{N}^{\alpha}=a^{*}b_{i}^{*}-N/N^{b}\left(F_{i3}^{b}/F_{33}^{b}V^{b}\right)/\mathcal{A},\quad i=1,2,3,\label{eq:Excesses  FV/F}
\end{equation}
where as before N refers to the total number of atoms inside the region
$ab^{*}c^{*}d$ spanned by the crossing vector $ab^{*}$ and $N^{b}$
is the number of atoms in the volume of single crystal defined by
lattice sites $a$, $b$, $c$, and $d$. For $i=3$, Eq.\ (\ref{eq:Excesses  FV/F})
recovers the well known expression for excess GB volume $[V]_{N}=(V-N\Omega)/A$.
Eq.\ (\ref{eq:Excesses  FV/F}) relates the three components of the
crossing vector to the excess properties of the GB.

We can now apply this equation to GB phases $\alpha$ and $\beta$
separately and evaluate the crossing vectors $\mathbf{A^{\prime}B^{\prime}}$
and $\mathbf{C^{\prime}D^{\prime}}$ in the Burgers circuit analysis:

\begin{equation}
A^{\prime}B_{i}^{\prime}=[VF_{i3}/F_{33}]_{N}^{\alpha}-\left(\Omega F_{i3}^{b,\alpha}/F_{33}^{b,\alpha}N^{\alpha}\right)/\mathcal{A},\quad i=1,2,3,\label{eq:A'B'_F_alpha}
\end{equation}

\begin{equation}
C^{\prime}D_{i}^{\prime}=[VF_{i3}/F_{33}]_{N}^{\beta}-\left(\Omega F_{i3}^{b,\beta}/F_{33}^{b,\beta}N^{\beta}\right)/\mathcal{A},\quad i=1,2,3.\label{eq:C'D'_F_beta}
\end{equation}
Notice that the bulk deformation gradients $F^{\ensuremath{b,\alpha}}$
and $F^{\ensuremath{b,\beta}}$ are not identical, as they depend
on how the lattice sites $AB$ and $CD$ were chosen. It is possible
to choose the same bulk reference state for both GB phases and use
the $F^{\alpha}$ and $F^{\beta}$ maps to predict the locations of
the sites $B^{\prime}$ and $C^{\prime}$. In this case, the A and
D corners of the Burgers circuit can still be selected arbitrarily,
but their counterparts in the upper grain are determined by the deformation
gradient $F^{b}$ and the number of atoms in the reference bulk crystal
or $V^{\prime}/\mathcal{A}^{\prime}$. Without loss of generality
we assume that the lattice sites in both bicrystals are chosen such
that they are related by the same $F^{b}$. Then, subtracting Eqs.\ (\ref{eq:A'B'_F_alpha})
and (\ref{eq:C'D'_F_beta}) we obtain the following expression for
the Burgers vector

\begin{equation}
b_{i}=\Delta[VF_{i3}/F_{33}]_{N}+\left(\Omega F_{i3}^{b}/F_{33}^{b}\Delta N\right)/\mathcal{A}+d_{i}^{sc},\quad i=1,2,3,\label{eq: BV_3components_F}
\end{equation}
where, as before, $\Delta N$ corresponds to the difference in the
number of atoms in bicrystals spanned by the crossing vectors, and
the DSC vector appears as a result of adding the lattice vectors $\mathbf{\mathbf{B^{\prime}C^{\prime}}}$
and $\mathbf{\mathbf{D^{\prime}A^{\prime}}}$. Eq.\ (\ref{eq: BV_3components_F})
shows that all possible Burgers vectors form a DSC lattice with the
origin shifted by a vector $\Delta[VF_{i3}/F_{33}]_{N}+\left(\Omega F_{i3}^{b}/F_{33}^{b}\Delta N\right)/\mathcal{A}$,
which components contain excess volume, excess shears and a term related
to the difference in the number of atoms. We can further reduce the
second term in Eq.\ (\ref{eq: BV_3components_F}) to $\left(\Omega F_{i3}^{b}/F_{33}^{b}\Delta N^{*}\right)/\mathcal{A}$
by subtracting out all DSC vectors, which is equivalent to selecting
one of the Burgers vectors closest to the origin of the shifted DSC
lattice, and obtain : 

\begin{equation}
b_{i}=\Delta[VF_{i3}/F_{33}]_{N}+(\Omega F_{i3}^{b}/F_{33}^{b}\Delta N^{*})/\mathit{\mathcal{A}}+d_{i}^{sc},\quad i=1,2,3.\label{eq:BV_3components_F_N*}
\end{equation}
Eq.\ (\ref{eq:BV_3components_F_N*}) is a vector form of Eq.\ (\ref{eq:b_normal_alytic_V_N_dSC})
derived previously for only one component $b_{3}$. The first term
in Eq.\ (\ref{eq:BV_3components_F_N*}) represents the contribution
to the Burgers vector from the difference in excess volumes and excess
shears, while the second is related to the the number of atoms required
to transform one GB phase into another. These two terms containing
properties specific to the two GB phases represent the non-DSC vector
by which the origin of the DSC lattice of all possible Burgers vectors
of the junction is shifted relative to zero.

As discussed previously,\citep{Frolov2012a,Frolov2012b} the excess
shear as an equilibrium property is not defined for GBs that move
under applied shear stress. As an example, consider symmetric tilt
GBs. When such a GB moves, one grain transforms into the other with
a different shape. Thus the deformation gradients in the two grains
are not the same. When such a boundary does not move, $\mathbf{F}^{b}$
can be assumed to be the same in both grains and a relation between
lattice sites across the GB has to be established to calculate the
formal $\bar{\mathbf{F}}^{\alpha}$. One way to establish this relation
is by following the bulk lattice sites during the GB creation procedure
such as the $\gamma$-surface approach. For example, Pond and Vitek
used this approach to track relative displacements of the grains and
calculated the Burgers vectors of partial GB dislocations formed by
identical GB structures corresponding to different grain translation
vectors.\citep{pond_periodic_1977_1,pond_periodic_1977_2} While this
procedure is straightforward if the boundary structures can be obtained
simply by translating and relaxing the adjacent crystals, it cannot
be applied if the adjacent GB phases are composed of different numbers
of atoms. 

On the other hand, even for boundaries that move by coupled motion,
such as symmetric tilt boundaries, it is straightforward to calculate
the excess shear component parallel to the tilt axis and use Eq.\ (\ref{eq:BV_3components_F_N*})
to predict $\mathbf{b}$ along that direction. A recent experimental
and modeling study demonstrated phases of {[}111{]} symmetric and
asymmetric tilt GBs in Cu that had different translations along the
tilt axis. According to Eq.\ (\ref{eq:BV_3components_F_N*}), GB
phase junctions of these boundaries have a screw component along the
{[}111{]} tilt axis. GBs with significant grain translations along
the tilt axis have been previously reported\citep{doi:10.1080/01418610208240038,PhysRevLett.70.449}
for other boundaries. The translations are typically on the order
of a half of the inter-planar distance. These translations result
in large excess shears and produce junctions that have a large screw
component parallel with the tilt axes, such as the one studied in
Ref.\ \citep{meiners_observations_2020}. Quantification of these
Burgers vectors using the described Burgers circuit analysis and Eq.\ (\ref{eq: BV_3components_F})
allows one to make predictions about the stability of metastable states
and explain the slow kinetics of GB phase transformations observed
in Ref.\ \citep{meiners_observations_2020}. 

\section{Burgers vectors of GB phase junctions in $\Sigma5$ symmetrical tilt
Cu boundaries\label{sec: MD structures analysis}}

\begin{figure}
\includegraphics[width=0.7\textwidth]{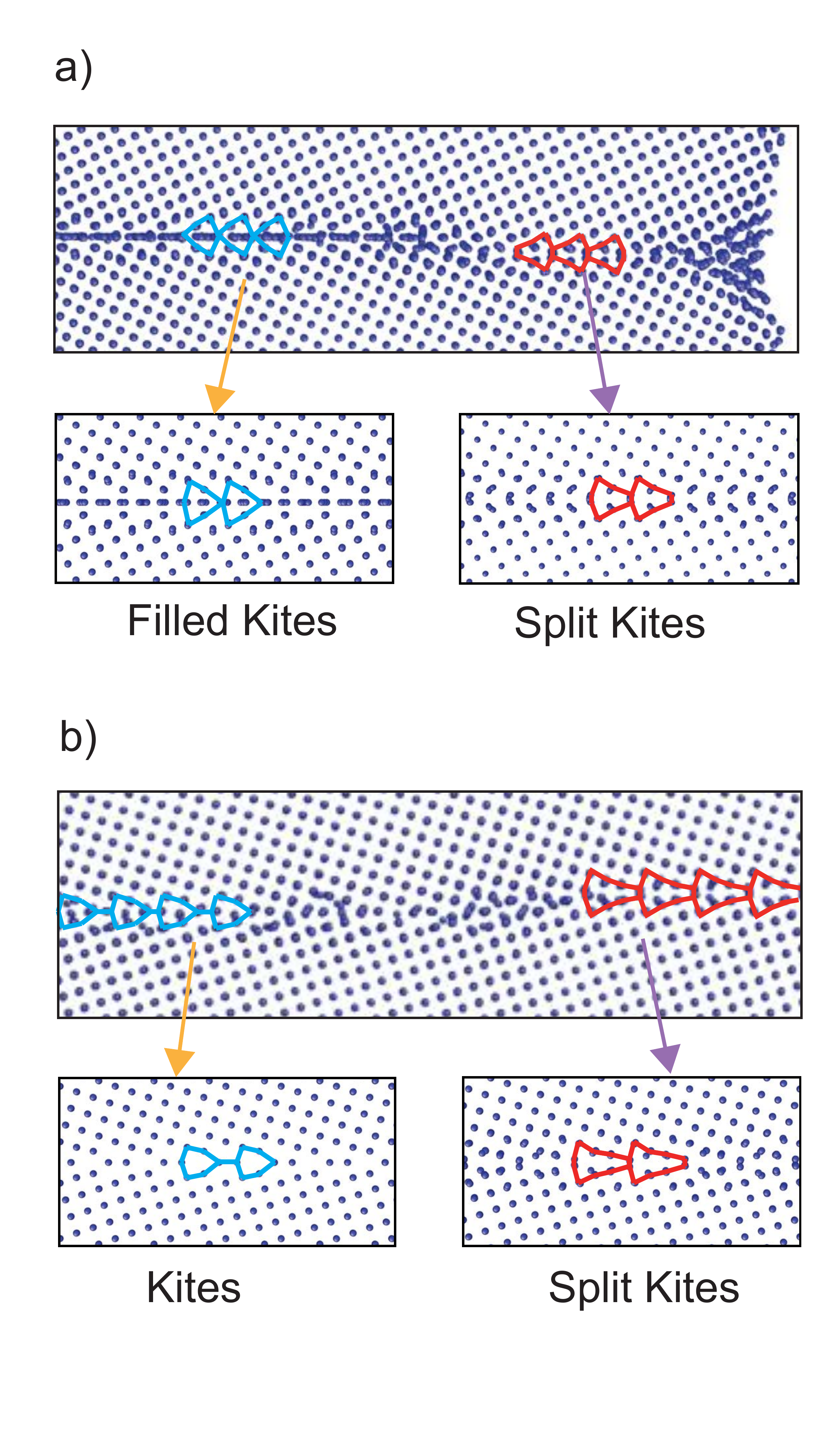}

\caption{Molecular dynamics simulations of structural transformations in a)
$\Sigma5(210)[001]$ and b) $\Sigma5(310)[001]$ symmetric tilt GBs
in Cu at T=800\ K from Ref.\ \citep{Frolov2013}. a) GB phase junction
formed by Filled Kites and Split Kites GB phases. b) GB phase junction
formed by Kites and Split Kites GB phases.The insets show zoomed-in
views of the corresponding GB structures computed at 0\ K.\label{fig:MD_310_210} }
\end{figure}

We now apply the methodology described above to analyze two specific
GB phase junctions in the $\Sigma5(310)[001]$ and $\Sigma5(210)[001]$
Cu boundaries\citep{Frolov2013} shown in Fig.\ \ref{fig:MD_310_210}.
First, we calculate the vectors $\mathbf{b}$ using the Burgers circuit
construction described in Sec.\ \ref{subsec:Closure-failure}. Then
we predict $\mathbf{b}$ using Eqs\ (\ref{eq:b_normal_alytic_V_N_dSC})
and (\ref{eq: BV_3components_F}) with the GB excess properties summarized
in Table\ \ref{tab:Properties of GB phases}, and compare the values
of $\mathbf{b}$ obtained by the two methods. 

\subsection{Analysis of the $\Sigma5(310)[001]$ GB}

\begin{figure}
\begin{centering}
\includegraphics[height=0.8\textheight]{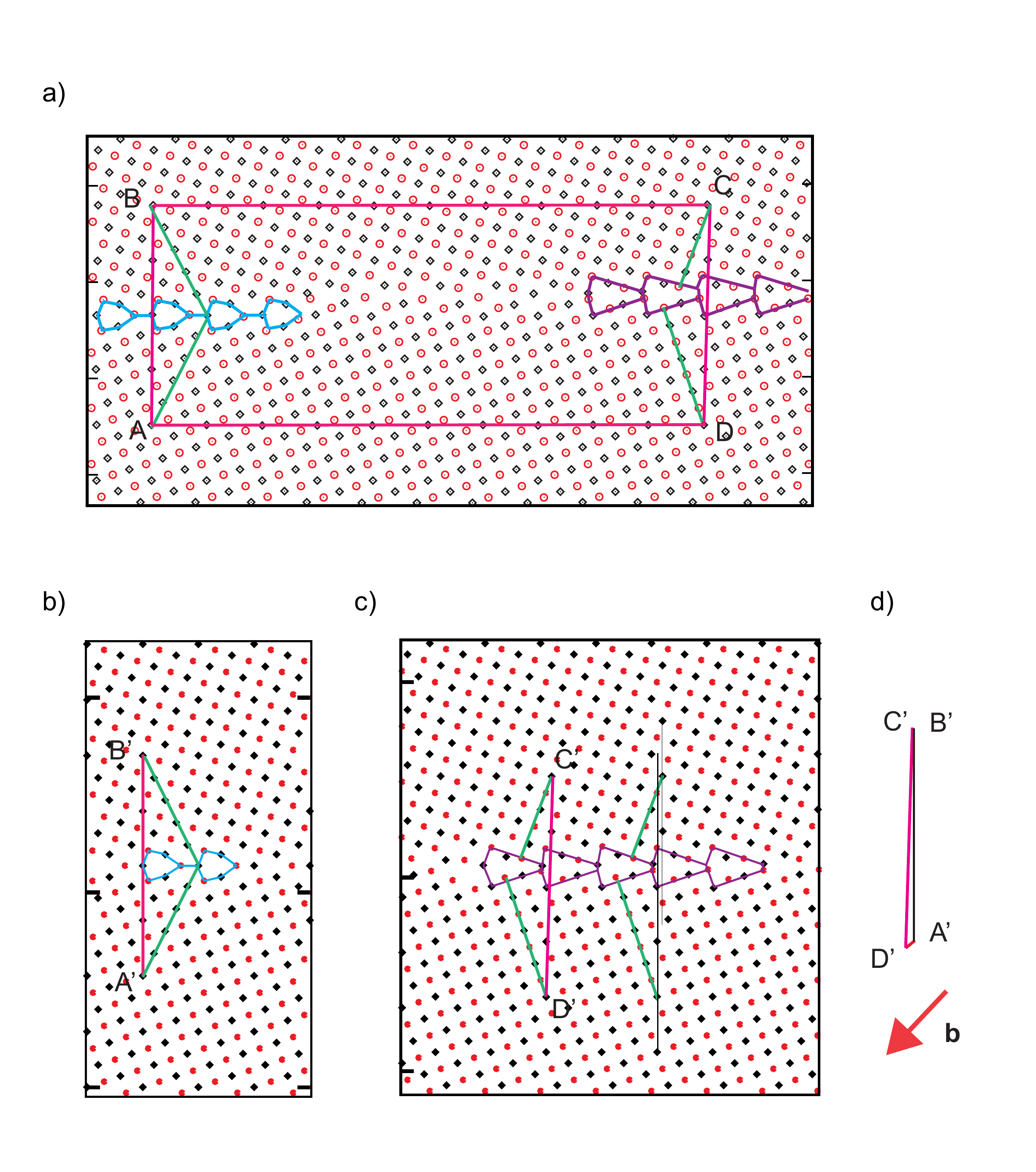}
\par\end{centering}
\caption{\label{fig:Circuit310}Calculation of the Burgers vector $\mathbf{b}$
of phase junction in the $\Sigma5(310)[001]$ GB using a closed circuit
ABCD. a) Deformed state containing the GB phase junction. The Kite
structure is on the left, while the Split Kite structure is on the
right. For convenience, both the $\mathbf{BC}$ and $\mathbf{DA}$
vectors are chosen to lie in (310) planes and have the same length
in the reference state. By this choice, their contribution to $\mathbf{b}$
is zero. b) and c) show bicrystals with Kite and Split Kite phases
in the reference state. To map the lattice sites $A$, $B$, $C$
and $D$ from the deformed state onto their positions $A^{\prime}$,
$B^{\prime}$, $C^{\prime}$ and $D^{\prime}$ in the reference state,
we follow (100) planes marked by green lines. In the Split Kite structure,
the lattice points $C^{\prime}$ and $D^{\prime}$ are offset by $d_{\{260\}}$
parallel to the interface, which is indicated by two vertical black
lines. d) The Burgers vector $\mathbf{b}$ is equal to the sum $\mathbf{A'B'}+\mathbf{C'D'}$.}
\end{figure}

Figure\ \ref{fig:Circuit310}a shows a closed circuit ABCD around
the GB phase junction in the $\Sigma5(310)[001]$ GB. For convenience,
we consider a slice parallel to the tilt axis containing only two
atomic planes. The atoms with different coordinates normal to the
page are colored in red and black. Vectors $\mathbf{AB}$ and $\mathbf{CD}$
cross the Kite and Split Kite GB structures, respectively. To simplify
the analysis, we have chosen the lattice sites A, B, C and D such
that the vectors $\mathbf{B'C'}$ and $\mathbf{D'A'}$ have the same
length equal to $10d_{\{310\}}=10(a/2)\sqrt{10}$ in the reference
state, where $a=3.615$ $\lyxmathsym{\AA}$ is the fcc lattice constant.\citep{Mishin01}
Since the two vectors have the same magnitude and opposite signs,
they cancel each other in Eq.\ (\ref{eq:bcircuit}) and do not contribute
to $\mathbf{\mathbf{b}}$. The reference bicrystals are shown in Fig.\ \ref{fig:Circuit310}b
and c. For these two simulation blocks, the boundary conditions are
periodic in the boundary plane and the stresses inside the bulk crystals
away from the boundary are zero. The exact simulation procedure is
not important as long as the structures of the boundaries match the
ones in the two GB phase state. In reality, the structures generated
at 0 K and those taken out from the simulation at T=800K were not
identical. The high temperature structures contain point defects and
may have a somewhat different arrangement of the atoms. Nevertheless,
the 0 K and finite temperature structures are very close. Table \ref{tab:Properties of GB phases}
summarizes the properties of the different GB phases calculated at
0~K.\citep{Frolov2013} In this work, we used both the 0 K and the
finite temperature structures to generate the reference bicrystals
and obtained the same result within the expected error of the calculation
method.

\begin{figure}
\begin{centering}
\includegraphics[width=0.9\textwidth]{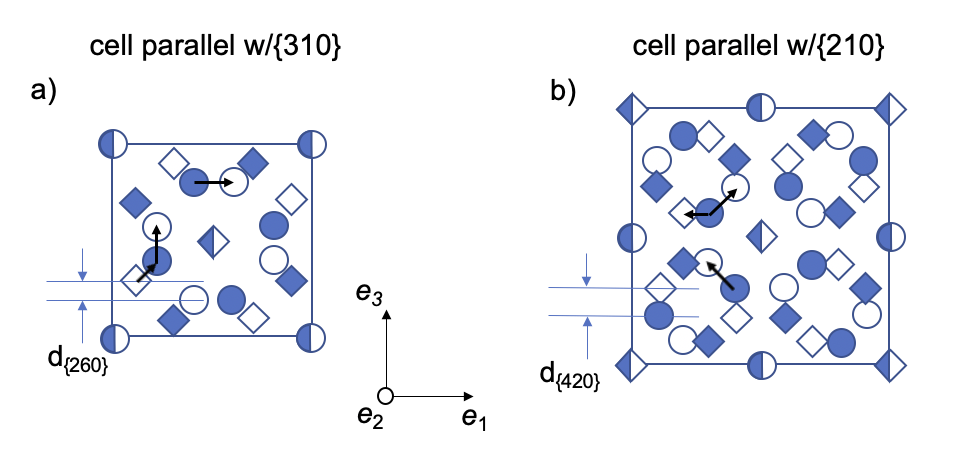}
\par\end{centering}
\caption{\label{fig:Dichrom}The dichromatic patterns for bicrystals with a)
$\Sigma5(310)[001]$ and b) $\Sigma5(210)[001]$ GBs. The filled and
open symbols distinguish the lattice sites belonging to the two different
grains. The lattice sites represented by diamonds are shifted relative
to the sites represented by circles by a/2 normal to the plane of
the figure. CSL in-plane edges are a) $a/2<310>=a\sqrt{10}/2$ and
b) $a<210>=a\sqrt{5}$, where a is the lattice parameter of fcc Cu
\citep{Mishin01}. $d_{\{260\}}=a/(2\sqrt{10})$ and $d_{\{420\}}=a/(2\sqrt{5})$
correspond to the distances between atomic planes inside the crystals
along the directions normal to the planes of the boundaries. Black
arrows indicate the vectors of the DSC lattice: a) $\mathbf{d_{1}^{SC}}=(a/\sqrt{10},0,0)$,
$\mathbf{d_{2}^{SC}}=(a/(2\sqrt{10)},-a/2,a/(2\sqrt{10)})$, $\mathbf{d_{3}^{SC}}=(0,0,a/\sqrt{10})$
and b) $\mathbf{d_{1}^{SC}}=(a/(2\sqrt{5}),0,a/(2\sqrt{5}))$, $\mathbf{d_{2}^{SC}}=(-a/(2\sqrt{5}),a/2,0\mathbf{)}$,
, $\mathbf{d_{3}^{SC}}=(-a/(2\sqrt{5}),0,a/(2\sqrt{5}))$. }
\end{figure}

To map the crossing vectors $\mathbf{AB}$ and $\mathbf{CD}$ in the
deformed state on the vectors $\mathbf{A'B'}$ and $\mathbf{C'D'}$
in the reference state, we follow the lattice planes in both crystals
as indicated by the green lines shown in Fig. \ref{fig:Circuit310}.
The exact choice of the guiding lines is not important as long as
they help to establish the relation between the lattice points A and
B on different sides of the GB. Performing a direct calculation of
the components of the crossing vectors in the reference state, we
obtained for the two bicrystals $A'B'_{3}=22.61$ $\lyxmathsym{\AA}$,
$A'B'_{1}=0$ $\lyxmathsym{\AA}$ and $C'D'_{3}=22.73$ $\lyxmathsym{\AA}$,
$C'D'_{1}=-d_{\{260\}}=-0.572$ $\lyxmathsym{\AA}$. The non-zero
$C'D'_{1}$ shown in Fig.\ \ref{fig:Circuit310}c indicates that
in the Split Kite structure the upper crystal is translated to the
right relative to the bottom crystal by the amount of $d_{\{260\}}=a/(2\sqrt{10})$.
At the same time, the Kite structure is symmetric with $A'B'_{1}=0$
. We can express the normal components of the crossing vectors in
terms of the bicrystal and GB contributions. $A'B'_{3}=22.61$$\,\lyxmathsym{\AA}=39d_{\{260\}}+[V]_{N}^{K}$
and $C'D'_{3}=22.73$ $\lyxmathsym{\AA}=39d_{\{260\}}+[V]_{N}^{SK}+0.38d_{\{260\}}$,
where $d_{\{260\}}=a/(2\sqrt{10})$ and represents the shortest distance
between two atomic planes in the crystal parallel to the GB plane.
The smallest DSC vector normal to the GB plane has the length $2d_{\{260\}}$
, as shown in Fig. \ref{fig:Dichrom}a, as a result even for simple
Kite structure $\mathbf{A'B'}-[V]_{N}^{K}\mathbf{n}_{GB}=(0,0,A'B'_{3}-[V]_{N}^{K})=(0,0,39d_{\{260\}})$
($\mathbf{n}_{GB}$ is the unit vector normal to the GB plane) is
not a DSC vector, which is not surprising because GBs allow for grain
translations parallel to the GB plane. The non-DSC part of the crossing
vector equals to $d_{260}$ reflects that translation and can be used
as a useful GB descriptor. While all crossing vectors form a DSC lattice,
the origin of this lattice is also shifted normal to the GB plane
by $d_{260}$. Notice that for this boundary $A'B'_{3}-[V]_{N}^{K}$
is equal to an integer number times the smallest normal component
of DSC equal to $d_{\{260\}}$. The Split Kite structure cannot be
obtained by joining two perfect half crystals and requires an insertion
or removal of a fraction of atoms less than one atomic plane. This
is reflected by the $0.38d_{\{260\}}$ terms in the expression for
$C'D'_{3}$.

Summing up the measured vectors of the circuit using Eq.\ (\ref{eq:bcircuit})
we obtain the components of the Burgers vector of the GB phase junction:
$b_{1}=-0.572\,\lyxmathsym{\AA}=-d_{\{260\}}$ $\lyxmathsym{\AA}$,
$b_{3}=-0.121$ $\lyxmathsym{\AA}$. Since the entire circuit was
located in one plane and there were no grain translations parallel
to the tilt axis, $b_{2}=0$ $\lyxmathsym{\AA}$. The negative value
of $b_{3}$ indicates that the bicrystal with the Split Kite structure
is effectively thicker than the bicrystal with the Kite structure.
This result may seem counterintuitive since the Split Kite has a smaller
excess volume then the Kite phase according to Table \ref{tab:Properties of GB phases}.
To explain this result, we express the calculated Burgers vector in
terms of DSC vectors and the excess GB properties. 

We now apply Eq.\ (\ref{eq:b_normal_analytic_alpha_beta}) to predict
the Burgers vector for the junction in the $\Sigma5(310)[001]$ GB.
The excess volumes and the numbers of atoms in the two GB phases,
Kites and Split Kites, can be found in Table\ \ref{tab:Properties of GB phases}.
TheGB areal number density is expressed as a fraction of the number
of atoms in one lattice plane parallel to the GB. The excess volume
of the Split Kite phase is smaller than that of the Kites, as a result
$\Delta[V]_{N}=[V]_{N}^{K}-[V]_{N}^{SK}=0.316\text{Å}-0.245\text{Å}=0.071\text{Å}$
is positive, while $\Omega\Delta N^{*}/A=\Omega(N^{K}-N^{SK})/\mathcal{A}=-0.38d_{\{260\}}=-0.21$
$\lyxmathsym{\AA}$ is negative. Summing up these two contributions
we obtain a negative value $b_{3}=-0.14\mathring{A}$, indicating
that the bicrystal with Split Kite structure is indeed effectively
thicker than the bicrystal with the Kite structure. The value of $b_{3}$
calculated using Eq.\ (\ref{eq:b_normal_analytic_alpha_beta}) also
matches the value obtained above using the Burgers circuit construction
within the numerical accuracy. This agreement suggests that the dislocation
content of this particular GB phase junction originates entirely from
the difference in excess properties of the two GB structures\textit{
i.e.}, from the difference in their excess volumes and the numbers
of atoms. Indeed, the number-of-atoms term $\Omega\Delta N^{*}/A=\Omega(N^{K}-N^{SK})/\mathcal{A}=-0.21$
$\lyxmathsym{\AA}$ is well defined: during the transformation this
exact amount of atoms per unit of area diffused from the open surface
and transformed the initial Kite phase to the Split Kite phase, as
was confirmed in Ref.\ \citep{Frolov2013}. This change in the number
of atoms can be easily evaluated by counting the total number of atoms
inside two regions on the two sides of the GB phase junction, containing
the two different GB phases with the same area. Such a calculation
was performed in the original study.\citep{Frolov2013} In other simulations
similar junctions were formed by inserting a controlled number of
atoms into the preexisting parent phase. 

The parallel component of the disconnection arises from the relative
shift of the grains parallel to the GB plane, which is different in
the two structures. For the given junction this difference is equal
to a DSC vector $b_{1}=-d_{\{260\}}=-0.57$ $\lyxmathsym{\AA}$. Note
that $b_{1}$ is smaller than the shortest DSC vector with the same
direction, with has the length $2d_{\{260\}}$. In general, the parallel
relative shift for a GB structure is not constrained to be a DSC vector.
The parallel components of the crossing vectors can be expressed as
a sum of DSC vector components and the excess shears at the boundary
as described by Eqs.\ \ref{eq: BV_3components_F}. As was discussed
above, symmetric tilt boundaries move under shear in that particular
direction and excess shear becomes ill-defined. For a stationary boundary,
this component of the Burgers vector can be formally interpreted to
have a contribution from the differences in the excess shears and
the numbers of atoms of the two GB phases as described by Eq.\ (\ref{eq:BV_3components_F_N*}). 

A GB phase junction can change its dislocation content by absorbing
or ejecting GB disconnections with Burgers vectors given by vectors
of the DSC lattice. We can consider such reactions and compare the
Burgers vector of the GB phase junction obtained in MD to other possible
valid vectors. The current Burgers vector in the coordinate frame
of the interface simulation is given by $\mathbf{b^{MD}}=(-d_{\{260\}},0,\Delta[V]_{N}-0.38d_{\{260\}})$,
while the primitive DSC lattice vectors are $\mathbf{d_{1}^{SC}}=(2d_{\{260\}},0,0)$,
$\mathbf{d_{2}^{SC}}=(d_{\{260\}},-a/2,d_{\{260\}})$ and $\mathbf{d_{3}^{SC}}=(0,0,2d_{\{260\}})$,
. Fig.\ \ref{fig:Dichrom}a shows the dichromatic pattern constructed
for this bicrystal as well as the primitive DSC lattice vectors.\citep{doi:10.1098/rsta.1979.0069,pond_bicrystallography_1983}
It is clear, that all possible disconnection reactions leave the magnitude
of the current Burgers vector at best unchanged. For example, consider
an absorption of $\mathbf{d_{1}^{SC}}$, which can glide in the boundary.
This dislocation reaction changes the direction of the GB phase junction
Burgers vector to $\mathbf{b^{MD}+d_{1}^{SC}}=(d_{\{260\}},0,\Delta[V]_{N}-0.38d_{\{260\}})$,
but not its magnitude. Other disconnection reactions increase the
magnitude of the Burgers vector and we conclude, that this junction
formed when the Split Kite structure absorbed extra atoms gives the
smallest Burgers vector possible for this GB. 

\begin{figure}
\raggedright{}\includegraphics[width=0.9\textwidth]{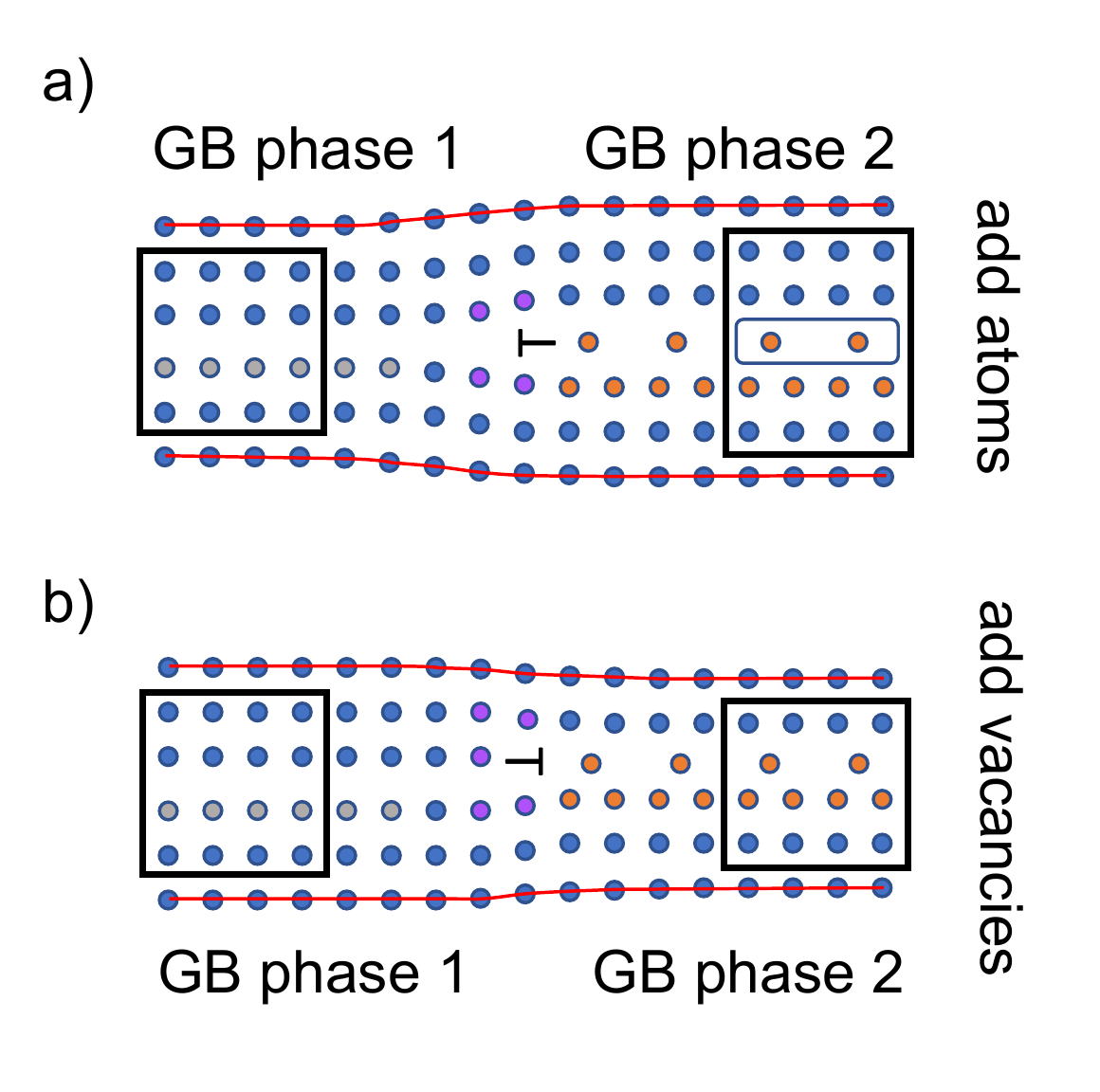}\caption{\label{fig:Schem_inter_vac}Schematic illustration of two GB phase
junctions with different Burgers vectors formed by the same GB phases.
For clarity, the upper and lower grains appear to have the same orientation
in this projection. The gray and orange lattice sites represent the
different core structures. The purple lattice sites indicate the GB
phase junctions. The black rectangles indicate the equivalent volumes
on the two sides of the junctions, they are bound by the same lattice
planes indicated by the red lines. a) GB phase 2, mimicking the Split
Kite phase, is formed by adding extra atoms to the GB phase 1 (Kite
phase). There two additional atoms in the region on the right, which
corresponds to +0.5 fraction of a bulk plane. b) GB phase 2 is formed
by adding the same fraction of vacancies to the GB phase 1. While
the GB phases are the same in a) and b) the junctions are different.
The defect absorption capacities $\Delta N'/A$ given by the difference
in the numbers of atoms in the equivalent volumes per unit are are
also different for the two junctions. }
\end{figure}

The analysis of other possible Burgers vectors can be used to explain
why the GB transformation in our simulation proceeded by absorption
of extra atoms and not vacancies. We can also predict possible Burgers
vectors for vacancy induced transformations. In the absence of mechanical
stresses, the primary driving force for the GB phase transformations
is the free energy difference between the Kite and Split Kite phases.
The Split Kite phase can be obtained from Kites by inserting a number
of atoms equal to 0.38 fraction atoms in a bulk plane parallel to
the boundary. An insertion or removal of a complete atomic plane (1.0
fraction) accompanied by a required grain translation restores the
original GB structure. Because of this periodicity, the Split Kite
phase can also be obtained from Kites by removing (1-0.38)=0.62 fraction
of atoms, \textit{i.e.,} by inserting this amount of vacancies. In
both transformations, we obtain a junction between the same phases:
Kites and Split Kites, but the Burgers vector of the junction is different.
A schematic illustration in Fig.\ \ref{fig:Schem_inter_vac} shows
how two different junctions between the same GB phases can be formed
by inserting extra atoms or vacancies into the same parent GB phase. 

Using the available GB properties, we can predict the smallest normal
component of the Burgers vector due to this hypothetical transformation
due to vacancies. Since the phases obtained are identical, the excess
volume contribution to $b_{3}$ is the same, $\Delta[V]_{N}=[V]_{N}^{K}-[V]_{N}^{SK}=0.071\text{Å}$.
The number of atoms term, on the other hand, has a different magnitude
and sign $\left(\Omega\Delta N^{*}/\mathcal{A}\right)^{vacancy}=\Omega(N^{K}-N^{SK})/\mathcal{A}=(1-0.38)d_{\{260\}}=0.35$
$\lyxmathsym{\AA}$. Summing up the two contributions we obtain $\mathbf{b}_{3}^{vacancy}=0.62d_{\{260\}}+\Delta[V]_{N}=0.35+0.071=0.42\:\mathring{A}$,
which is larger than the normal component obtained in the MD simulations.
One of the smallest possible Burgers vector with this normal component
is $\mathbf{b^{MD}+d_{2}^{SC}}=(0,-a/2,\Delta[V]_{N}+0.62d_{\{260\}})$,
has a much larger magnitude due to the non-zero component along the
tilt axis. The large energetic penalty due to nucleation of this dislocation
makes the transformation by absorption of $0.62$ fraction of a plane
of vacancies less likely. 

Another valid Burgers vector consistent with the vacancy absorption
mechanism is $\mathbf{b^{MD}}+\mathbf{d_{3}^{SC}}=(-d_{\{260\}},0,\Delta[V]_{N}+1.62d_{\{260\}})$.
It has a larger normal component compare to $0.62d_{\{260\}}+\Delta[V]_{N}$,
but a much smaller magnitude of the total Burgers vector. Instead
of absorbing 0.62 fraction of vacancies to nucleate the Split Kite
phase, this mechanism requites absorbing that fraction of vacancies
plus a complete lattice plane of vacancies. The difference between
the two Burgers vectors due to atoms and vacancies absorption is $\mathbf{d_{3}^{SC}}=(0,0,2d_{\{260\}})$,
with $2d_{\{260\}}$ corresponding two atomic planes, not one. Thus,
the presented analysis makes a prediction about the difference in
the transformation of Kite structure into Split Kite by atom and vacancy
absorption mechanisms. The vacancy induced transformation requires
about three times larger amount of point defects to be absorbed per
unit of the transformed GB area than the interstitial induced transformation.
Even this smallest Burgers vector consistent with the vacancy induced
transformation mechanism has a magnitude larger than $\mathbf{b^{MD}}$,
suggesting that in the absence of mechanical stresses such a transformation
by vacancy absorption is less energetically favorable compared to
the transformation by the absorption of atoms, which we observed in
our MD simulations. When mechanical stresses are applied additional
driving forces appear that can influence the transformation. 

\subsection{Analysis of the $\Sigma5(210)[001]$ GB}

\begin{figure}
\raggedright{}\includegraphics[width=0.9\textwidth]{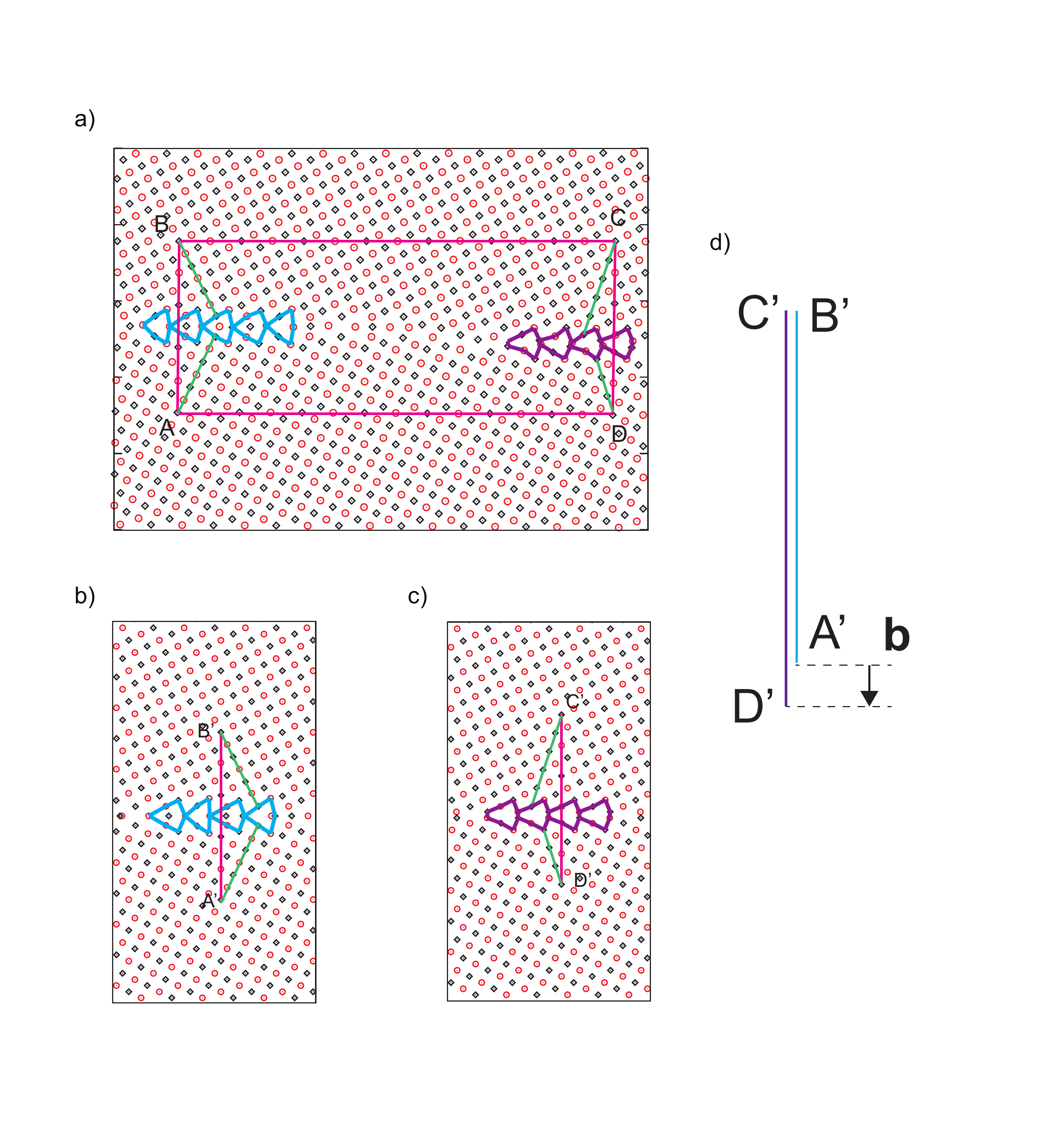}\caption{\label{fig:210_BV}Calculation of the Burgers vector $\mathbf{b}$
of phase junction in the $\Sigma5(210)[001]$ GB using a closed circuit
ABCD. a) Deformed state containing the GB phase junction. The Filled
Kite structure is on the left, while the Split Kite structure is on
the right. For convenience, the $\mathbf{BC}$ and $\mathbf{DA}$
vectors are chosen to lie in (210) planes and are have the same length
in the reference state. By this choice, their contribution to $\mathbf{b}$
is zero. b) and c) show bicrystals with Filled Kite and Split Kite
phases in the reference state. To map the lattice sites $A$, $B$,
$C$ and $D$ from the deformed state onto their positions $A^{\prime}$,
$B^{\prime}$, $C^{\prime}$ and $D^{\prime}$ in the reference state,
we followed lattice planes indicated by the green lines. d) The Burgers
vector $\mathbf{b}$ is equal to the sum $\mathbf{A'B}'+\mathbf{C'D'}$.}
\end{figure}

$\Sigma5(210)[001]$ is another symmetric tilt GB studied in Ref.\ \citep{Frolov2013}.
This boundary shows a first-order transition between Filled-Kite and
Split-Kite phases. Generation of both GB phases requires insertion
or removal of atoms. Unlike the Kite phase, they cannot be obtained
using the $\gamma$-surface approach, \textit{i.e.}, by simply translating
the two grains laterally with respect to each other and parallel with
the GB plane. Figure\ \ref{fig:210_BV}a shows the slice of the structure
containing two atomic planes with atoms colored in red and black according
to their position along the tilt axis. As before, we construct a closed
circuit ABCD and identify the crossing vectors $\mathbf{A'B'}$ and
$\mathbf{C'D'}$ in reference bicrystals as shown in Fig. \ref{fig:210_BV}.
The calculated components of these vectors in the reference state
are $A'B'_{3}=22.0$ $\lyxmathsym{\AA}$ , $A'B'_{1}=0$ $\lyxmathsym{\AA}$
and $C'D'_{3}=22.36$ $\lyxmathsym{\AA}$, $C'D'_{1}=0$ $\lyxmathsym{\AA}$.
Both Filled-Kite and Split-Kite phases required insertion or removal
of atoms and we can express the calculated components as $A'B'_{3}=27d_{\{420\}}+[V]_{N}^{FK}-1/7d_{\{420\}}$
and $C'D'_{3}=27d_{\{420\}}+[V]_{N}^{SK}+7/15d_{\{420\}}$, where
$d_{\{420\}}=a/(2\sqrt{5})=0.81$ $\lyxmathsym{\AA}$ and represents
the shortest distance between two atomic planes in the crystal parallel
to the GB plane. The smallest DSC vector strictly normal to the GB
plane has the length $2d_{\{420\}}$, while $d_{\{420\}}$ correspond
to a smallest normal component of a vector on the DSC lattice, as
shown in Fig.\ \ref{fig:Dichrom}b.

Using Eq.\ (\ref{eq:bcircuit}) and the measured crossing vectors,
we obtain the following components for the Burgers vector: $b_{3}=-0.36$
$\lyxmathsym{\AA}$ and $b_{1}=0$ $\lyxmathsym{\AA}$. Similarly
to the $\Sigma5(310)[001]$ boundary, the entire circuit is located
in the same plane in the reference state, so the component $b_{2}$
parallel to the tilt axis is also zero. The Burgers vector components
calculated directly using the closed circuit approach can also be
interpreted in terms of the excess properties of the Filled Kite and
Split Kite phases. According to Table\ \ref{tab:Properties of GB phases},
the difference in the excess volumes is $\Delta[V]_{N}=[V]_{N}^{FK}-[V]_{N}^{SK}=0.301\text{Å}-0.172\text{Å}=0.129\text{Å}$,
while the difference in the number of atoms gives $\Omega\Delta N^{*}/\mathcal{A}=\Omega(N^{FK}-N^{SK})/\mathcal{A}=-(7/15+1/7)d_{\{420\}}=0.60\cdot0.81\text{Å}=-0.49$
$\lyxmathsym{\AA}$. Adding these two terms we obtained the normal
component of the Burgers vector predicted by the Eq.\ (\ref{eq:b_normal_analytic_alpha_beta})
to be $b_{3}=\Delta[V]_{N}+\Delta N^{*}\cdot\Omega/\mathcal{A}=-0.36$$\lyxmathsym{\AA}$.
Since the relative tangential translation vectors are zero for both
bicrystals, both $b_{1}$ and $b_{2}$ are zero. These numbers again
match well the components calculated using the Burgers circuit analysis.

Similar to the first GB phase junction, here, we can also conclude
that the obtained Burgers vector is the smallest possible. Indeed,
the current Burgers vector is given by $\mathbf{b^{MD}}=(0,0,\Delta[V]_{N}-0.6d_{\{420\}})$,
while the primitive DSC lattice vectors are $\mathbf{d_{1}^{SC}}=(d_{\{420\}},0,d_{\{420\}})$,
$\mathbf{d_{2}^{SC}}=(-d_{\{420\}},a/2,0)$, $\mathbf{d_{3}^{SC}}=(-d_{\{420\}},0,d_{\{420\}})$.
Fig. \ref{fig:Dichrom}b shows the dichromatic pattern constructed
for this bicrystal as well as the primitive DSC lattice vectors. Adding
of any of these DSC vectors to $\mathbf{b^{MD}}$ will not decrease
the magnitude of the resultant Burgers vector of the GB phase junction.

Similarly to the analysis of the $\Sigma5(310)[001]$ GB, here, we
can also consider a hypothetical transformation in which the Split-Kite
phase of this boundary grows via absorption of vacancies instead of
atoms. The excess volume component of $b_{3}$ remains again the same
$\Delta[V]_{N}=[V]_{N}^{FK}-[V]_{N}^{SK}=0.129\text{Å}$, while the
second contribution from atoms becomes $\left(\Omega\Delta N^{*}/\mathcal{A}\right)^{vacancy}=\Omega(N^{FK}-N^{SK})/\mathcal{A}=(1-7/15-1/7)d_{\{420\}}=0.4d_{\{420\}}=0.32$
$\lyxmathsym{\AA}$. Summing up these two contributions we obtain
the smallest normal component $b_{3}^{vacancy}=0.129+0.32=0.55\:\mathring{A}$.
A possible valid Burgers vector for such a transformation could be
for example $\mathbf{b^{MD}}+\mathbf{d_{1}^{SC}}=(d_{\{420\}},0,\Delta[V]_{N}+0.4d_{\{420\}})$,
which is also the smallest Burgers vector consistent with the vacancy
induced transformation. Notice that for this boundary the difference
in the normal components of the burgers vector is $d_{\{420\}}$ which
corresponds to one atomic plane. As a result, there is no significant
difference in the amount of absorbed point defects by both mechanisms:
0.6 fraction of a plane of atoms is absorbed in one case, and 0.4
fraction of a plane of vacancies in the other. 

In MD simulations, both $\Sigma5$ boundaries transformed to the Split
Kite phase by absorption of extra atoms, not vacancies. Our analysis
indicates that, when extra atoms are absorbed the two contributions
to the Burgers vector from the difference in the excess volumes and
the numbers of atoms have opposite signs resulting in a smaller Burgers
vector of the GB phase junction, making this transformation more energetically
favorable when no external mechanical stresses are applied. 

\section{Discussion and Conclusions}

In this work, we have analyzed the dislocation content of GB phase
junctions. Like dislocations, these line defects generate long-range
elastic fields and can interact with other defects such as regular
GB disconnections, dislocations, surfaces and precipitates. During
GB phase nucleation, the elastic interaction between GB phase junctions
and their strain energy contributes to the nucleation barrier. Understanding
the Burgers vectors of these defects is necessary to describe these
interactions and to quantify the nucleation barriers during GB phase
transformations. In this study, we have described a general Burgers
circuit approach that allows one to calculate Burgers vectors of junctions
formed by different GB structures composed of different numbers of
atoms. We also derived expressions that relate the components of the
Burgers vector to the differences in the properties of GB phases,
including excess volume, excess shears and the numbers of atoms $\Delta N^{*}$
required for the GB phase transformation. We showed that, differently
from regular GB disconnections, the Burgers vectors of GB phase junctions
are not DSC vectors. While all allowed Burgers vectors of a GB phase
junction form a DSC lattice, the origin of this lattice is shifted
by a non-DSC vector determined by the differences in the mentioned
GB properties and $\Delta N^{*}$. It has been recognized by prior
studies\citep{pond_periodic_1977_1,pond_periodic_1977_2} that the
difference between the grain translation vectors creates GB dislocations
when structures with different translation vectors coexist on the
the same plane. Pond and Vitek simulated partial GB dislocations formed
by identical GB structures with different relative grain translations
and defined the Burgers vector of these dislocations as the difference
between their translation vectors.\citep{pond_periodic_1977_1,pond_periodic_1977_2}
GB dislocations formed by different GB structures with different excess
properties and numbers of atoms have not been analyzed. It has also
been suggested that the difference in excess volumes of different
GB structures coexisting on the same plane contributes to the normal
component of the Burgers vector.\citep{pond_periodic_1977_1,pond_periodic_1977_2}
In this work we have shown that, when two different GB phases are
composed of different number of atoms, the normal component of the
Burgers vector is not equal to the difference in the excess volumes.
The difference in the numbers of atoms required for the GB phase transformation
also contributes to $\mathbf{b}$. 

We have applied this analysis to GB phase junctions modeled in the
$\Sigma5(210)[001]$ and $\Sigma5(310)[001]$ symmetric tilt GBs in
Cu. In both boundaries, these junctions are formed between two GB
phases with different structures and different numbers of atoms. The
Burgers vectors were calculated using two separate approaches. In
the first one, we used a straightforward Burgers circuit construction,
which characterizes the $\mathbf{b}$ components. In the second approach,
we used known values of excess properties of the studied GB phases
to predict the smallest components of the Burgers vectors normal to
the GB plane. The difference in the numbers of atoms was calculated
after the transformation took place. For both GB phase junctions studied,
the magnitudes of the Burgers vectors were found to be the smallest
possible and their normal components matched the ones predicted from
the known GB properties. The obtained Burgers vectors had two non-zero
components and one zero component parallel to the tilt axis. The normal
component of the Burgers vector was not equal to the difference in
the excess volumes and contained a second contribution due to the
difference in the numbers of atoms $\Delta N^{*}$ required for the
GB phase transformation. For the $\Sigma5(310)[001]$ boundary, this
later contribution was larger than the difference in the excess volumes
and even had an opposite sign. For both junctions studied, the contribution
to $\mathbf{b}$ from the difference in the numbers of atoms $\Delta N^{*}$
is significant and cannot be neglected. In our analysis we considered
absorption or ejection of additional disconnections with Burgers vectors
dictated by the DSC lattice and concluded that these reactions cannot
further reduce the calculated $\mathbf{b}$. We also showed that some
larger predicted Burgers vectors corresponded to GB phase transformations
that proceed by absorption of vacancies. This analysis could explain
why both GBs transformed by absorbing extra atoms and not vacancies.

The multiplicity of the possible burgers vectors of GB phase junctions
formed between the same GB phases has important implications for GB
phase equilibrium and the kinetics of GB phase transformations. In
elemental fluid systems, interfacial phases in equilibrium have the
same Gibbs free energy, which means that their excess Helmholtz free
energy difference is balanced by the $-P\Delta[V]_{N}$ term.\citep{Gibbs}
The later term represents the mechanical work per unit area done by
the pressure $P$ during the transformation. This condition is analogous
to the co-existence conditions of bulk phases under pressure. Since
the excess volume difference is the only interface property that couples
to external stress this coexistence state is unique and is defined
by the excess properties of the interfacial phases. In solid systems,
the generalized analog of the $-P\Delta[V]_{N}$ term is the work
per unit area of the PK force that acts on the GB phase junction.
When GB phases are in contact with particle reservoirs such as open
surfaces that enable the potential change in the number of GB atoms
(or the system is closed but $\Delta N^{*}=0$), the equilibrium is
established when the difference in the excess GB Helmholtz free energies
is balanced by the PK force on the GB phase junction. Since the PK
force depends on the Burgers vector of the phase junction, the equilibrium
coexistence between the same GB phases can be established at different
temperatures and stresses depending on the Burgers vector of the junction.
Similarly, the driving force for the GB phase transformation for a
given temperature and stress is not determined by the GB phases alone
and also depends on the Burgers vector of the junction. For example,
in this work, we considered different junctions between the same GB
phases formed by the insertion of vacancies and interstitials and
showed that the normal components of their Burgers vectors have opposite
signs. When the same stress normal to the GB plane is applied the
PK force will drive these two junctions in opposite directions. Moreover,
the $\Delta N^{*}$ contribution to the Burgers vector may change
the PK force in the way that normal compression no longer favors the
GB phase with the smallest excess volume, as usually expected. These
considerations demonstrate that the dislocation nature of GB phase
junctions makes GB phase transformations richer than similar transformations
at interfaces in fluid systems.

The investigation of dislocation properties of GB phase junctions
have implications for our understanding of GB phase transitions. At
present, the role of elastic interactions in the kinetics of GB phase
transformations is not well-understood. At the same time, there is
growing modeling evidence suggesting that such interactions could
be important.\citep{Frolov2013,meiners_observations_2020} Recent
experimental and modeling study suggested that barriers associated
with the motion of the GB phase junction could be responsible for
the slow kinetics of such transformations and could stabilize metastable
GB states. Modeling studies also showed that nucleation at surface
triple junctions is much more effective than homogenous nucleation
even when sources of atoms are not required.\citep{meiners_observations_2020}
Nucleation models that incorporate elastic interactions have been
recently developed for regular GB disconnections to describe GB migration
and interactions with triple junctions.\citep{han_grain-boundary_2018,thomas_reconciling_2017,Thomas8756}
The present analysis suggests that similar nucleations models should
be developed for GB phase transformations to gain further insight
into their energetics and kinetics. 

\section*{Acknowledgment }

This work was performed under the auspices of the U.S. Department
of Energy (DOE) by Lawrence Livermore National Laboratory under contract
DE-AC52-07NA27344. T.F. was funded by the Laboratory Directed Research
and Development Program at Lawrence Livermore National Laboratory
under Project Tracking Code number 19-ERD-026. DLM was funded by the
U.S. Department of Energy (DOE), Office of Science, Basic Energy Sciences
(BES), Materials Science and Engineering Division (MSE). Sandia National
Laboratories is a multi-mission laboratory managed and operated by
National Technology \& Engineering Solutions of Sandia, LLC, a wholly
owned subsidiary of Honeywell International Inc., for the U.S. Department
of Energy\textquoteright s National Nuclear Security Administration
under contract DE-NA0003525. This paper describes objective technical
results and analysis. Any subjective views or opinions that might
be expressed in the paper do not necessarily represent the views of
the U.S. Department of Energy or the United States Government. M.A.
acknowledges support from the Office of Naval Research under grant
number N0014-19-1-2376. The authors are grateful to Yuri Mishin and
David Olmsted for valuable discussions. T.F. is grateful to Ian Winter
for stimulating discussions. 


\begin{thebibliography}{10}

\bibitem{Cantwell20141}
P.~R. Cantwell, M.~Tang, S.~J. Dillon, J.~Luo, G.~S. Rohrer, and M.~P. Harmer,
\newblock Acta Materialia {\bf 62}, 1  (2014).

\bibitem{Frolov2013}
T.~Frolov, D.~L. Olmsted, M.~Asta, and Y.~Mishin,
\newblock Nat. Commun. {\bf 4}, 1899 (2013).

\bibitem{Gibbs}
J.~W. Gibbs,
\newblock {\em The Scientific Papers of J. Willard Gibbs}, volume~1,
\newblock Longmans-Green, London, 1906.

\bibitem{Frolov:2015ab}
T.~Frolov and Y.~Mishin,
\newblock J. Chem. Phys. {\bf 143}, 044706 (2015).

\bibitem{PhysRevB.95.155444}
R.~Freitas, T.~Frolov, and M.~Asta,
\newblock Phys. Rev. B {\bf 95}, 155444 (2017).

\bibitem{Divinski2012}
S.~V. Divinski, H.~Edelhoff, and S.~Prokofjev,
\newblock Phys. Rev. B {\bf 85}, 144104 (2012).

\bibitem{PhysRevLett.59.2887}
K.~L. Merkle and D.~J. Smith,
\newblock Phys. Rev. Lett. {\bf 59}, 2887 (1987).

\bibitem{krause_review_2019}
A.~R. Krause, P.~R. Cantwell, C.~J. Marvel, C.~Compson, J.~M. Rickman, and
  M.~P. Harmer,
\newblock Journal of the American Ceramic Society {\bf 102}, 778 (2019).

\bibitem{Rickman2016225}
J.~Rickman and J.~Luo,
\newblock Curr. Opin. Solid State Mater. Sci. {\bf 20}, 225  (2016).

\bibitem{doi:10.1080/095008399177020}
E.~Rabkin, C.~Minkwitz, C.~Herzig, and L.~Klinger,
\newblock Philosophical Magazine Letters {\bf 79}, 409 (1999).

\bibitem{Rohrer2016231}
G.~S. Rohrer,
\newblock Curr. Opin. Solid State Mater. Sci. {\bf 20}, 231  (2016).

\bibitem{rupert_role_2016}
T.~J. Rupert,
\newblock Current Opinion in Solid State and Materials Science {\bf 20}, 257
  (2016).

\bibitem{Dillon20076208}
S.~J. Dillon, M.~Tang, W.~C. Carter, and M.~P. Harmer,
\newblock Acta Mater. {\bf 55}, 6208 (2007).

\bibitem{OBrien2018}
C.~J. O'Brien, C.~M. Barr, P.~M. Price, K.~Hattar, and S.~M. Foiles,
\newblock Journal of Materials Science {\bf 53}, 2911 (2018).

\bibitem{ABDELJAWAD2017528}
F.~Abdeljawad, P.~Lu, N.~Argibay, B.~G. Clark, B.~L. Boyce, and S.~M. Foiles,
\newblock Acta Materialia {\bf 126}, 528  (2017).

\bibitem{rajeshwari_k_grain_2020}
S.~Rajeshwari~K., S.~Sankaran, K.~C. Hari~Kumar, H.~Rosner, M.~Peterlechner,
  V.~A. Esin, S.~Divinski, and G.~Wilde,
\newblock Acta Materialia {\bf 195}, 501 (2020).

\bibitem{glienke_grain_2020}
M.~Glienke, M.~Vaidya, K.~Gururaj, L.~Daum, B.~Tas, L.~Rogal, K.~G. Pradeep,
  S.~V. Divinski, and G.~Wilde,
\newblock Acta Materialia {\bf 195}, 304 (2020).

\bibitem{HIRTH2013749}
J.~Hirth, R.~Pond, R.~Hoagland, X.-Y. Liu, and J.~Wang,
\newblock Progress in Materials Science {\bf 58}, 749  (2013).

\bibitem{Hirth96}
J.~P. Hirth and R.~C. Pond,
\newblock Acta Mater. {\bf 44}, 4749 (1996).

\bibitem{Medlin2017383}
D.~Medlin, K.~Hattar, J.~Zimmerman, F.~Abdeljawad, and S.~Foiles,
\newblock Acta Materialia {\bf 124}, 383  (2017).

\bibitem{medlin_accommodation_2003}
D.~L. Medlin, D.~Cohen, and R.~C. Pond,
\newblock Philosophical Magazine Letters {\bf 83}, 223 (2003).

\bibitem{medlin_accommodation_2006}
D.~L. Medlin, D.~Cohen, R.~C. Pond, A.~Serra, J.~A. Brown, and Y.~Mishin,
\newblock Microscopy and Microanalysis {\bf 12}, 888 (2006).

\bibitem{Pond03a}
R.~C. Pond and S.~Celotto,
\newblock Int. Mater. Rev. {\bf 48}, 225 (2003).

\bibitem{hirth_disconnections_2016}
J.~P. Hirth, J.~Wang, and C.~N. Tome,
\newblock Progress in Materials Science {\bf 83}, 417 (2016).

\bibitem{rajabzadeh_role_2014}
A.~Rajabzadeh, F.~Mompiou, S.~Lartigue-Korinek, N.~Combe, M.~Legros, and D.~A.
  Molodov,
\newblock Acta Materialia {\bf 77}, 223 (2014).

\bibitem{zhu_situ_2019}
Q.~Zhu, G.~Cao, J.~Wang, C.~Deng, J.~Li, Z.~Zhang, and S.~X. Mao,
\newblock Nature Communications {\bf 10}, 156 (2019),
\newblock Number: 1 Publisher: Nature Publishing Group.

\bibitem{meiners_observations_2020}
T.~Meiners, T.~Frolov, R.~E. Rudd, G.~Dehm, and C.~H. Liebscher,
\newblock Nature {\bf 579}, 375 (2020).

\bibitem{doi:10.1080/01418618308243118}
P.~W. Tasker and D.~M. Duffy,
\newblock Philos. Mag. A {\bf 47}, L45 (1983).

\bibitem{doi:10.1080/01418618608242811}
D.~M. Duffy and P.~W. Tasker,
\newblock Philos. Mag. A {\bf 53}, 113 (1986).

\bibitem{DUFFY84a}
D.~M. Duffy and P.~W. Tasker,
\newblock J. Am. Ceram. Soc {\bf 67}, 176 (1984).

\bibitem{Phillpot1992}
S.~R. Phillpot and J.~M. Rickman,
\newblock The Journal of Chemical Physics {\bf 97}, 2651 (1992).

\bibitem{Phillpot1994}
S.~R. Phillpot,
\newblock Phys. Rev. B {\bf 49}, 7639 (1994).

\bibitem{Alfthan06}
S.~\mbox{von Alfthan}, P.~D. Haynes, K.~Kashi, and A.~P. Sutton,
\newblock Phys. Rev. Lett. {\bf 96}, 055505 (2006).

\bibitem{Alfthan07}
S.~\mbox{von Alfthan}, K.~Kaski, and A.~P. Sutton,
\newblock Phys. Rev. {\rm B} {\bf 76}, 245317 (2007).

\bibitem{Chua:2010uq}
A.~L.~S. Chua, N.~A. Benedek, L.~Chen, M.~W. Finnis, and A.~P. Sutton,
\newblock Nat Mater {\bf 9}, 418 (2010).

\bibitem{Demkowicz2015}
W.~Yu and M.~Demkowicz,
\newblock Journal of Materials Science {\bf 50}, 4047 (2015).

\bibitem{Cahn79}
J.~W. Cahn,
\newblock Thermodynamics of solid and fluid surfaces,
\newblock in {\em Interface Segregation}, edited by W.~C. Johnson and J.~M.
  Blackely, chapter~1, page~3, American Society of Metals, Metals Park, OH,
  1979.

\bibitem{Frolov2012a}
T.~Frolov and Y.~Mishin,
\newblock Phys. Rev. B {\bf 85}, 224106 (2012).

\bibitem{Frolov2012b}
T.~Frolov and Y.~Mishin,
\newblock Phys. Rev. B {\bf 85}, 224107 (2012).

\bibitem{Hirth}
J.~P. Hirth and J.~Lothe,
\newblock {\em Theory of Dislocations},
\newblock Wiley, New York, 2 edition, 1982.

\bibitem{Note1}
In atomistic simulations this is always possible at least in principle. The
  reference GB phases can be generated separately or carved out from the
  deformed state and relaxed in a proper way so that the reference crossing
  vectors could be calculated. In experiment, the analysis cannot be completed
  based on a single image of the deformed state. However, GB structures
  sufficiently far away from the junction can be approximated as reference
  states. In general, a full three-dimensional structure is necessary to
  determine all three components of $\protect \mathbf {b}$.

\bibitem{Han:2016aa}
J.~Han, V.~Vitek, and D.~J. Srolovitz,
\newblock Acta Mater. {\bf 104}, 259 (2016).

\bibitem{Zhu2018}
Q.~Zhu, A.~Samanta, B.~Li, R.~E. Rudd, and T.~Frolov,
\newblock Nat. Commun. {\bf 9}, 467 (2018).

\bibitem{banadaki_efficient_2018}
A.~D. Banadaki, M.~A. Tschopp, and S.~Patala,
\newblock Computational Materials Science {\bf 155}, 466 (2018).

\bibitem{gao_interface_2019}
B.~Gao, P.~Gao, S.~Lu, J.~Lv, Y.~Wang, and Y.~Ma,
\newblock Science Bulletin {\bf 64}, 301 (2019).

\bibitem{yang_grain_2020}
C.~Yang, M.~Zhang, and L.~Qi,
\newblock Computational Materials Science {\bf 184}, 109812 (2020).

\bibitem{Robin74}
P.~Y. Robin,
\newblock Am. Miner {\bf 59}, 1286 (1974).

\bibitem{dingreville_interfacial_2008}
R.~Dingreville and J.~Qu,
\newblock Journal of the Mechanics and Physics of Solids {\bf 56}, 1944 (2008).

\bibitem{Larche_Cahn_78}
F.~C. Larche and J.~W. Cahn,
\newblock Acta Metall. {\bf 26}, 1579 (1978).

\bibitem{doi:10.1063/1.448644}
W.~W. Mullins and R.~F. Sekerka,
\newblock The Journal of Chemical Physics {\bf 82}, 5192 (1985).

\bibitem{pond_periodic_1977_1}
R.~C. Pond, V.~Vitek, and P.~B. Hirsch,
\newblock Proceedings of the Royal Society of London. A. Mathematical and
  Physical Sciences {\bf 357}, 453 (1977).

\bibitem{pond_periodic_1977_2}
R.~C. Pond,
\newblock Proceedings of the Royal Society of London. Series A, Mathematical
  and Physical Sciences {\bf 357}, 471 (1977).

\bibitem{doi:10.1080/01418610208240038}
G.~H. Campbell, M.~Kumar, W.~E. King, J.~Belak, J.~A. Moriarty, and S.~M.
  Foiles,
\newblock Philosophical Magazine A {\bf 82}, 1573 (2002).

\bibitem{PhysRevLett.70.449}
G.~H. Campbell, S.~M. Foiles, P.~Gumbsch, M.~R\"uhle, and W.~E. King,
\newblock Phys. Rev. Lett. {\bf 70}, 449 (1993).

\bibitem{Mishin01}
Y.~Mishin, M.~J. Mehl, D.~A. Papaconstantopoulos, A.~F. Voter, and J.~D. Kress,
\newblock Phys. Rev. B {\bf 63}, 224106 (2001).

\bibitem{doi:10.1098/rsta.1979.0069}
R.~C. Pond, W.~Bollmann, and F.~C. Frank,
\newblock Philosophical Transactions of the Royal Society of London. Series A,
  Mathematical and Physical Sciences {\bf 292}, 449 (1979).

\bibitem{pond_bicrystallography_1983}
R.~C. Pond and D.~S. Vlachavas,
\newblock Proceedings of the Royal Society of London. Series A, Mathematical
  and Physical Sciences {\bf 386}, 95 (1983),
\newblock Publisher: The Royal Society.

\bibitem{han_grain-boundary_2018}
J.~Han, S.~L. Thomas, and D.~J. Srolovitz,
\newblock Progress in Materials Science {\bf 98}, 386 (2018).

\bibitem{thomas_reconciling_2017}
S.~L. Thomas, K.~Chen, J.~Han, P.~K. Purohit, and D.~J. Srolovitz,
\newblock Nature Communications {\bf 8}, 1 (2017).

\bibitem{Thomas8756}
S.~L. Thomas, C.~Wei, J.~Han, Y.~Xiang, and D.~J. Srolovitz,
\newblock Proceedings of the National Academy of Sciences {\bf 116}, 8756
  (2019).

\end{thebibliography}

\begin{table}
\begin{raggedright}
\begin{tabular}{|c|c|c|c|}
\hline 
Structure & $\Delta N^{*}$ relative to Kite phase & $[U]_{N}$, J/m$^{2}$ & $[V]_{N}$, Å\tabularnewline
\hline 
\multicolumn{4}{|c|}{$\Sigma5(310)[001]$}\tabularnewline
\hline 
Kites (0 K) & $0$ & 0.9047 & 0.316\tabularnewline
\hline 
Split kites (0 K) & $2/5$  & 0.911 & 0.233\tabularnewline
\hline 
Split kites (MD) & $0.37$ & 0.920 & 0.245\tabularnewline
\hline 
\multicolumn{4}{|c|}{$\Sigma5(210)[001]$}\tabularnewline
\hline 
Kites (0 K) & $0$ & 0.951 & 0.322\tabularnewline
\hline 
Split kites (0 K) & $7/15$  & 0.936 & 0.172\tabularnewline
\hline 
Split kites (MD) & $0.46$ & 0.98 & 0.23\tabularnewline
\hline 
Filled kites (0 K) & $6/7$ & 0.953 & 0.301\tabularnewline
\hline 
\end{tabular}
\par\end{raggedright}
\caption{\textbf{Excess properties of different GB phases calculated in Ref.}\ \textbf{\citep{Frolov2013}
including numbers of atoms $\Delta N^{*}$ relative to Kite phase
expressed as a fraction of atoms in a bulk plane parallel to the GB,
excess energy and volume. $\Sigma5(310)$} and \textbf{$\Sigma5(210)$}
GBs in Cu modeled with the EAM potential \citep{Mishin01}. The energies
indicate the ground states at 0 K. The zero fractions of a plane for
Kite structures indicate that these GB structures can be created by
joining two perfect half-crystals, while the non-zero fractions indicate
that extra atoms have to be inserted or removed to generate the other
GB structures. The fractions of the inserted atoms are calculated
relative to the number of atoms in one atomic plane parallel to the
GB. \label{tab:Properties of GB phases}}
\end{table}

\end{document}